\documentclass[conference]{IEEEtran}

\usepackage{algpseudocode}
\usepackage{cite}
\usepackage{amsmath,amssymb,amsfonts}
\usepackage{textcomp}
\usepackage{xcolor}
\usepackage{microtype}
\usepackage{balance}
\usepackage{leading}
\usepackage{tabularx}
\usepackage{textcomp}
\usepackage{multirow}
\usepackage{framed}
\usepackage{listings}
\usepackage{float}
\usepackage{enumitem}
\usepackage{comment}
\usepackage{cleveref}
\usepackage[T1]{fontenc}
\usepackage[utf8]{inputenc}
\usepackage{subfig}
\usepackage{graphicx}

\usepackage[lined,linesnumbered,ruled]{algorithm2e}

\usepackage{graphicx}
\usepackage[detect-all]{siunitx}
\begin{document}

% \title{Time-Cost Tradeoffs in Cloud-based Distributed Machine Learning}
\title{Scavenger: A Cloud Service For Optimizing Cost and Performance of ML Training}
%\author{Paper \# 7214}

\author{\IEEEauthorblockN{Sahil Tyagi}
\IEEEauthorblockA{Indiana University Bloomington, USA \\
styagi@iu.edu}
\and
\IEEEauthorblockN{Prateek Sharma}
\IEEEauthorblockA{Indiana University Bloomington, USA \\
prateeks@iu.edu}
}

% Middlware: https://middleware-conf.github.io/2022/call-for-research-papers/
% 12 pages + refs. acm sigplan 10pt

% Ccgrid: 10 pages + refs. IEEE. Double blind https://ccgrid2023.iisc.ac.in/call-for-papers/ 

\maketitle

\begin{abstract}
Cloud computing platforms can provide the computational resources required for training large machine learning models such as deep neural networks.
While the pay-as-you-go nature of cloud virtual machines (VMs) makes it easy to spin-up large clusters for training models, it can also lead to ballooning costs.
The 100s of virtual machine sizes provided by cloud platforms also makes it extremely challenging to select the ``right'' cloud cluster configuration for training.
Furthermore, the training time and cost of distributed model training is highly sensitive to the cluster configurations, and presents a large and complex tradeoff-space. 

In this paper, we develop principled and practical techniques for optimizing the training time and cost of distributed ML model training on the cloud. 
Our key insight is that both the parallel and statistical efficiency must be considered when selecting the optimum job configuration parameters such as the number of workers and the batch size.
By combining  conventional parallel scaling concepts and new insights into SGD noise, we develop  models for estimating the time and cost on different cluster configurations.
Using the repetitive nature of training and our performance models, our Scavenger cloud service can search for optimum cloud configurations in a black-box, online manner. 
Our approach reduces training times by $2\times$ and costs by more than 50\%. 
Compared to an oracle-based approach, our performance models are accurate to within 2\% such that the search imposes an overhead of just 10\%.

%%% Local Variables:
%%% mode: latex
%%% TeX-master: "paper"
%%% End:

\end{abstract}

\section{Introduction}
\label{sec:intro}
% Motivation
The discovery of improved machine learning (ML) models has resulted in great advances in computer vision, language and speech processing, scientific computing, and many other areas.
These advances are primarily driven by increasingly computationally intensive models, such as deep neural networks (DNNs), being ``trained'' on large data sets. 
The ready availability of computing resources is a key enabler of machine learning, and cloud platforms can easily provide these resources. 
%Training and discovering these models requires a large amount of computing resources (several teraflops over several hours), which cloud platforms can easily provide. 
%
%Concurrently, cloud computing has emerged as the de-facto paradigm for providing computing and storage resources to applications, including and especially machine learning due to its large computing requirements. 
%The efficient use of cloud platforms can significantly decrease the cost of training models, expand and accelerate the adoption of AI and data-driven techniques, and strengthen the integration of machine learning into the computational ecosystem. 
However, current ML techniques and systems are ill-suited for making effective and efficient use of cloud resources, i.e., are not cloud-native.

ML models are often trained on large clusters of cloud virtual machines, but this often leads to prohibitive costs, because ML training techniques and frameworks like TensorFlow and PyTorch are oblivious to cost. 
Moreover, cloud platforms offer 100s of different virtual machine sizes and configurations with different cost/performance tradeoffs, making it extremely challenging to select the ``right'' type and quantity of cloud resources. 
Training large ML models on the cloud is thus often performed on sub-optimally configured cloud resources, leading to cost overruns, slow performance, and underutilized resources. 

These challenges also exist when optimizing the resource allocation for conventional distributed applications (such as map-reduce data processing) on the cloud~\cite{alipourfard2017cherrypick}.
However, model training also has other unique execution and synchronization characteristics and a large array of configuration knobs (such as number of workers and the batch size) which have significant impact on performance and resource efficiency.

In this paper, we present Scavenger, a service for optimizing the cloud training cost and time for ML models. 
Scavenger is a model-agnostic, black-box, fully online service built using TensorFlow, and searches for good configurations for distributed model training jobs. 
We use a performance-model guided search across a multi-dimensional configuration space to find the pareto-optimal configurations based on user preferences and constraints.
In its search for the best configuration, Scavenger  horizontally scales a training job by adding/removing workers, and vertically scales it by changing the batch size.

As a key first step towards understanding and optimizing training time and costs, we develop a new phenomenological performance model for data-parallel distributed model training. 
Its is phenomenological in the sense that our prediction accuracy improves with the degree of exploration, which is based on the type of search performed in the search space (\S\; \ref{sec:design}).
Our model uses both conventional parallel scaling concepts such as synchronization overheads, as well as fundamental performance artifacts of Stochastic Gradient Descent (SGD) based optimization. 
Unlike in classical parallel applications, we find that computation performed by parallel workers doesn't always compose because of the stochastic nature of the gradient computation. 
This \emph{statistical inefficiency} reduces the rate of the job's forward-progress, and imposes its own tradeoff on time and cost which also depends on the cluster configuration. 
We measure and consider this statistical inefficiency by using SGD noise (variance of gradients), and show how it can be used as a general \emph{scaling indicator}. 
%The job's forward progress can be reduced because of this statistical inefficiency, which leads to higher training times and costs for 
%
%We integrate  SGD noise (variance of gradients)  into our performance model, and show how it can be computed and used as a general \emph{scaling indicator} for statistical efficiency. 

Scavenger is a fully managed model training service requiring minimal user intervention, prior knowledge, or offline profiling. 
We use the repetitive and iterative nature of model training to briefly profile the job on different configurations and learn its performance profile by using the scaling indicators.
We minimize the overhead of this exploration and search phase by using lightweight model checkpointing, and obtain the cost and time tradeoff curves for different combinations of workers and batch sizes. 
The performance model is then used to run the reminder of the job on the ``best'' configuration based on user preferences and constraints.

Our profiling-based strategy of building the performance model is optimized to reduce the search cost.
We can build a full performance profile of an ML model by profiling on only a small subset of configurations.
We accomplish this by leveraging our phenomenological first-principles performance models that can be interpolated using linear regression---thus requiring only a \emph{partial search}. 
Since Scavenger is a cloud service, it also leverages repeated training of similar models (e.g., part of hyperparamter or neural architecture search), and reuses its learned performance model, to completely eliminate the exploration phase and search costs. 
Surprisingly, we find that the SGD noise can serve as model-agnostic scaling indicator, and even a ``universal'' average model can estimate performance with reasonable accuracy without any exploration or pilot jobs.

To the best of our knowledge, Scavenger is the first work which can optimize both cost and time in a fully online manner.
We build on recent work for SGD noise based scaling such as~\cite{johnson2020adascale, mccandlish2018empirical, mai_kungfu_2020, qiao2021pollux}, and use it for simple intuitive phenomenological models. 
By considering both the parallel and statistical efficiency, we are able to accurately predict the training time of a wide range of DNN models with minimal search overhead.  

Scavenger is an open-source library built on top of TensorFlow, and provides a practical, online, black-box, model-agnostic service for addressing the crucial problem of cost and performance optimization of distributed machine learning in the cloud.
In addition to the practical significance, we make the following research contributions: 
\begin{enumerate}
\item We provide a thorough empirical investigation of the cost and time tradeoffs in distributed ML model training, and show how parallel and statistical efficiency influence the performance. 
\item We show how the variance in gradients results in SGD noise, and how it can serve as a reliable scaling indicator for elastic horizontal and vertical scaling. 
\item We develop new models for predicting the performance for deep neural networks, which consider both parallel and statistical efficiency, and the aforementioned SGD noise. Our models predict training time and cost for different job configurations (number of workers and batch size), and construct full tradeoff curves and pareto frontiers, with very high accuracy of more than 98\%. 
\item Our models enable us to search for the optimum job and cluster configuration in a model-agnostic and online manner, and minimize various combinations of cost and time. Our techniques can find the ``right'' cloud configuration and reduce training time by more than $2\times$ compared to naive configurations. 

%\item We show how the SGD noise scale can be efficiently computed, and examine its efficacy in dynamic cluster environments. We develop an interpolation based approach that integrates with the training cluster size and the global batch size.
%\item Elastic training is challenging. Scavenger uses checkpoint-restart technique to resume training with updated workers and batch-size. The system works in a black-box, online manner and does not require offline profiling.
%\item We develop an online algorithm that reacts to resource availability changes on a cluster-wide level, which also scales the batch size based on the scalability indicator (SGD noise).
\end{enumerate}

\section{Background and Challenges}
\label{sec:bg}
In this section, we describe the performance tradeoffs faced by distributed ML training. 
These observations and insights guide our performance model presented in the next section. 

\begin{figure}
  \centering 
  \includegraphics[width=0.3\textwidth]{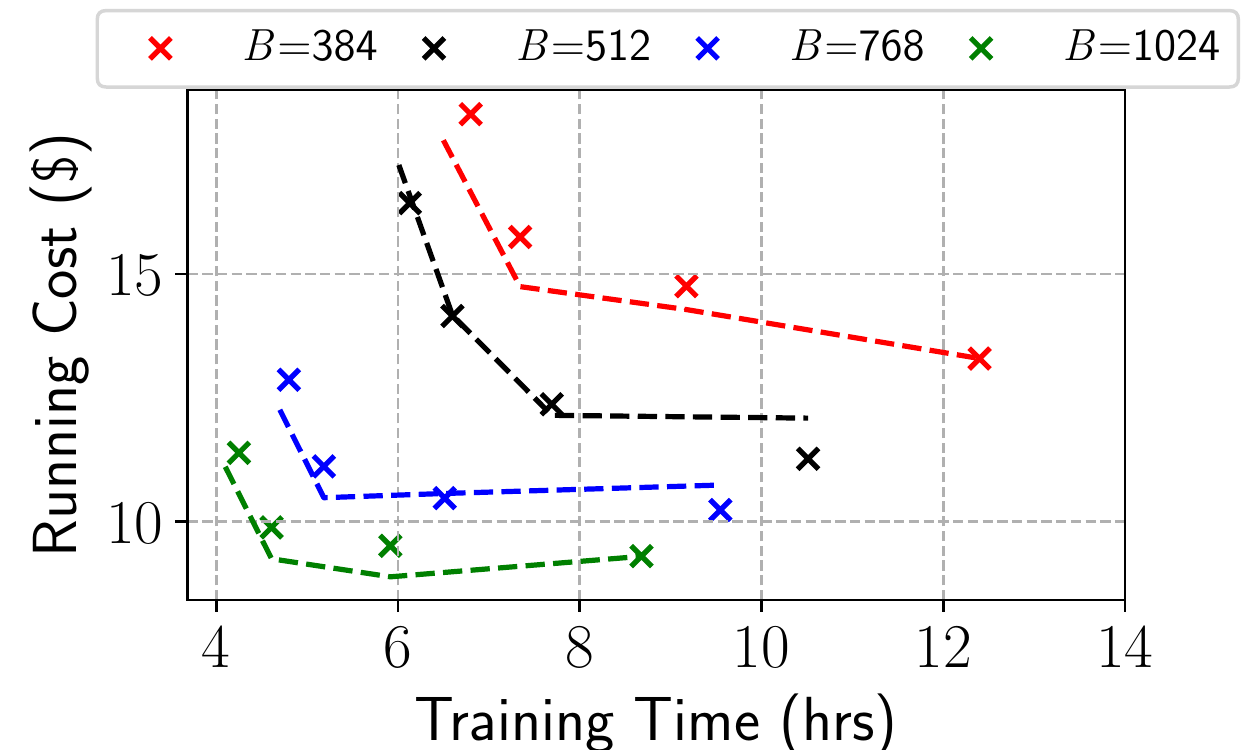}
  \caption{Running cost and time for different batch sizes and workers for ResNet18 training. Each point along a tradeoff-curve represents 20, 16, 12, 8 workers respectively. Dashed line shows our model prediction.}
  \label{fig:tradeoff-example}
  \vspace*{-0.5cm}
\end{figure}

\subsection{Distributed ML Training}

Distributed training entails learning the model parameters (or weights) of a model over an input training dataset.
A model trains in an iterative-convergent process to minimize a loss function over the dataset by using optimization techniques such as Stochastic Gradient Descent (SGD)~\cite{sgd} and Mini-Batch Gradient Descent~\cite{ruder2017overview} or Full Gradient Descent.

Since ML training is highly compute intensive, parallelizing it using computational accelerators such as GPUs and TPUs, and through distributed training, is vital~\cite{torsten_demystifying_2018,mayer_scalable_2019}.
A common parallelization approach is \emph{data-parallelism}, where training is launched on multiple workers, and each worker learns and updates the model parameters by processing a small batch of the training data~\cite{dean2012large} at each iteration.

After each iteration, the gradient updates from all workers are aggregated via all-reduce operations to compute the averaged gradients, update model parameters and synchronize the new parameters among the workers~\cite{li_scaling_2014}.
A popular and widely successful data-parallel training approach is the parameter server strategy, where the workers compute the gradients, and parameter servers aggregate and average the gradients from all workers after every iteration and update the model weights. 
Training a popular image recognition model like ResNet ~\cite{szegedy2017inception,he2015deep} or an attention-based language model like Transformer~\cite{vaswani2017attention} typically require thousands of iterations until the model's error converges to the desired low training loss. 

Concretely, the training process iteratively computes the model parameters over $K$ workers, each processing a mini-batch of $b$ at iteration $t$ and computing the gradient $\nabla f(\mathbf{x}_{k, t})$. The update rule for the model parameters $\mathbf{x}$ is given by:

\begin{equation}
  \mathbf{x}_{t+1} = \mathbf{x}_{t} - \eta \dfrac{1}{K} \dfrac{1}{b} \sum_{k=1}^{k=K}{\nabla f(\mathbf{x}_{k, t})},
  %\vspace*{\subsecspace}
    \label{eq:sgd}
\end{equation}
where $\eta$ is the learning rate, one of the hyperparameters of the model that is found through empirical search techniques.
The global batch size is $B = b K$, and is a crucial job parameter.

\begin{figure*}
  \centering
  \subfloat[ResNet18 \label{fig:cluster-scaling}]
  {\includegraphics[width=0.25\textwidth]{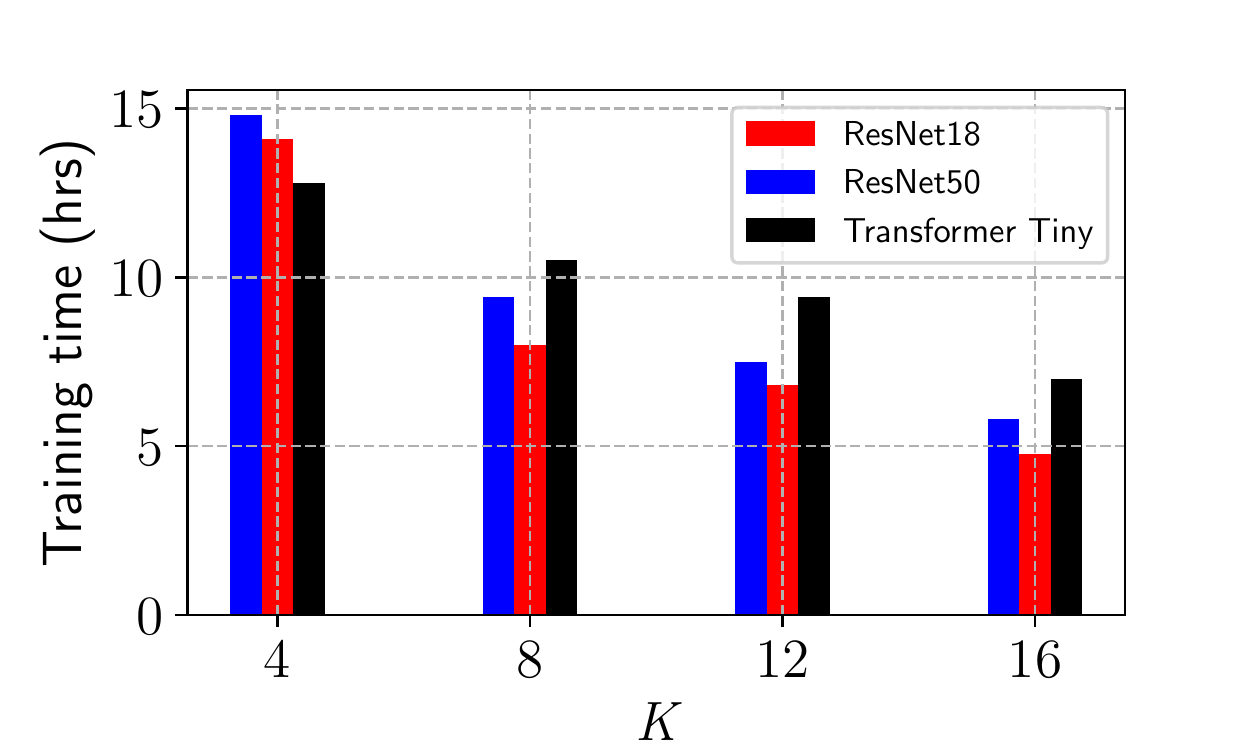}}	
  \subfloat[Transformer \label{fig:memory-limit}]
  {\includegraphics[width=0.25\textwidth]{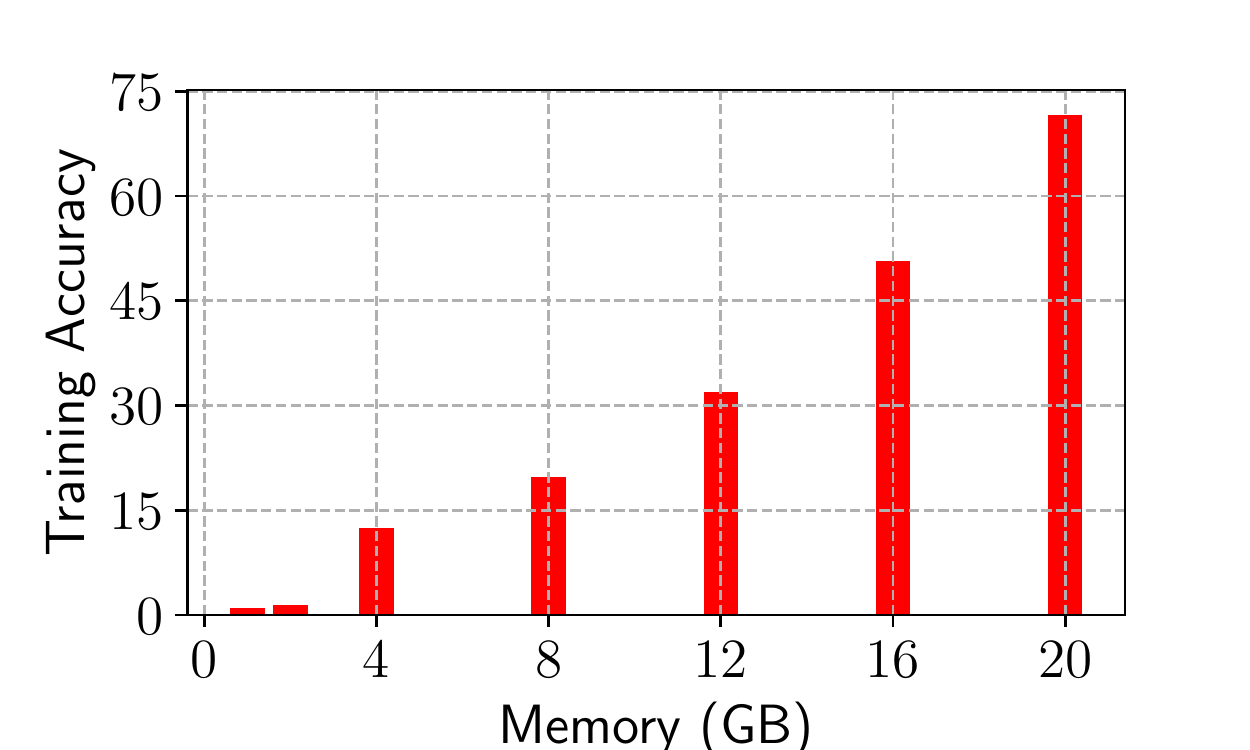}}
  \subfloat[ResNet18 \label{fig:loss-steps-examples}]
  {\includegraphics[width=0.25\textwidth]{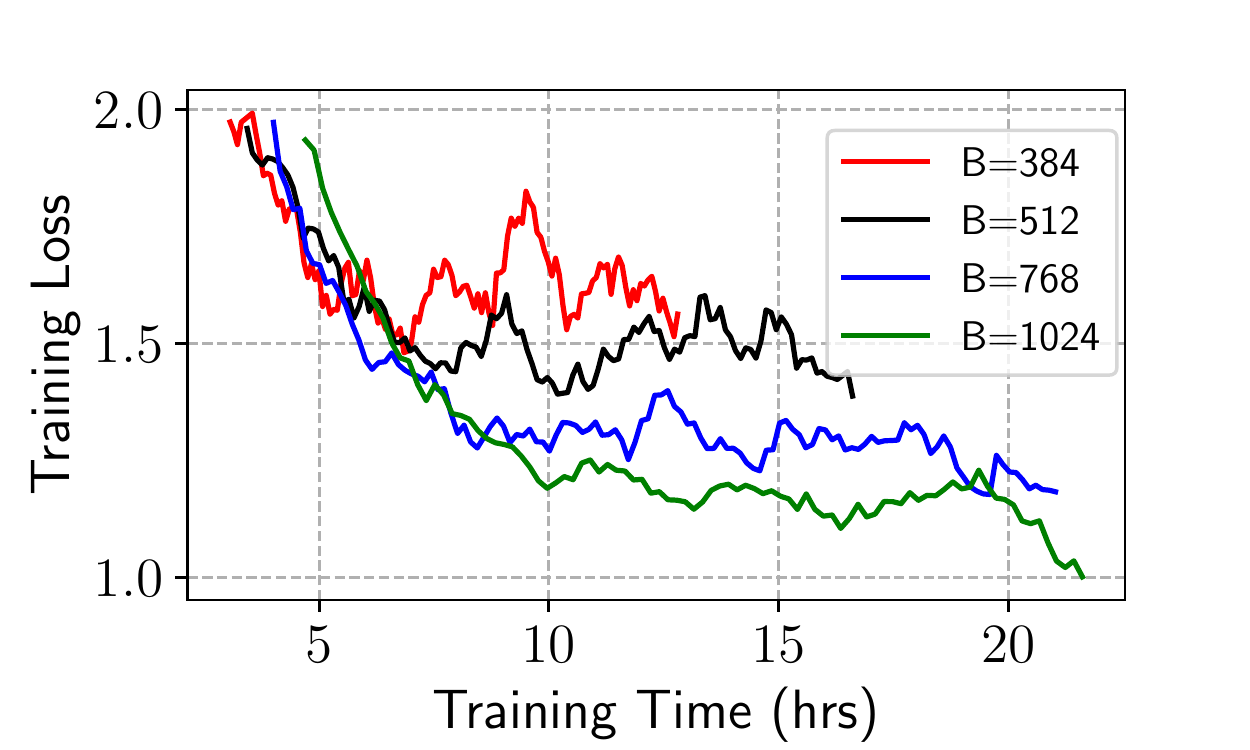}}
  \caption{(a) Improvement in training time as cluster size and capacity scales. ResNet18, ResNet50 and Transformer Tiny are run upto 80\% accuracy on CIFAR-10,90\% on CIFAR-100 and 15.0 BLEU on WMT14 dataset. (b) Accuracy reached by Transformer Base before job fails due to memory constraints. (c) Training time taken by ResNet18 on CIFAR-10 to converge.  \textit{Larger $\mathit{B}$ are more time-efficient to achieve the same model quality.}}
  \label{fig:hscale}
    \vspace*{-0.5cm}
\end{figure*}

\noindent \textbf{Elasticity.}
Distributed training is resource \emph{elastic}, which means that the models can be trained on different cluster sizes and configurations, which can also be changed during runtime (i.e., during model training).
Training can be horizontally scaled by adjusting the number of workers (i.e., changing $K$ in Equation~\ref{eq:sgd}), and vertically scaled by increasing the mini-batch size $b$ on each worker.
ML frameworks such as TensorFlow and PyTorch also support model checkpointing, and thus we can adjust the horizontal and vertical scaling dynamically by checkpointing the model state and resuming the training on a different cluster configuration.

This elasticity makes distributed training a good fit for clouds, since we can easily scale the cluster by adding/removing VMs, and changing the underlying VM size to increase the batch size and intra-worker parallelism.
Scavenger makes use of this elasticity in its search for the ideal cloud configuration. 
However, distributed training has complex and incompletely-understood performance tradeoffs~\cite{gupta_model_2016} that are affected by the various SGD parameters (such as $K, b$). 
Simply running more workers and increasing batch size has diminishing returns, as we can see from Figure~\ref{fig:tradeoff-example}, which shows the running time and cost for training the ResNet18 model.
Each point corresponds to a different number of workers for each batch size. 
We can see that there are diminishing returns, and thus it is not obvious which cluster configuration is the ``best''.

% which can also happen to be dynamic.
% This elasticity is well matched with the cloud: since different types of VMs can be added.
% %
% Distributed training has complex and ill-understood trade-offs between parallel efficiency, synchronization overheads, and the quality of the trained model.
% Like most parallel applications, adding more workers and increasing the parallelism has diminishing returns due to the periodic parameter synchronization (Amdahl's law~\cite{yavits2013effect}). However, additional scaling considerations apply.
% %
% Two key scaling knobs. The first is the number of workers, $K$, which can be achieved by provisioning and using more virtual machines (VMs).
% This increases the parallelism.

\subsection{Horizontal Scaling: Adding Workers}
\label{subsec:amdahl}

\begin{figure*}
		\centering
		%\hspace{1.5cm}
		%\centering
		\subfloat[ResNet18 \label{fig:resnet18-computesynctime}]
  		{\includegraphics[width=0.25\textwidth]{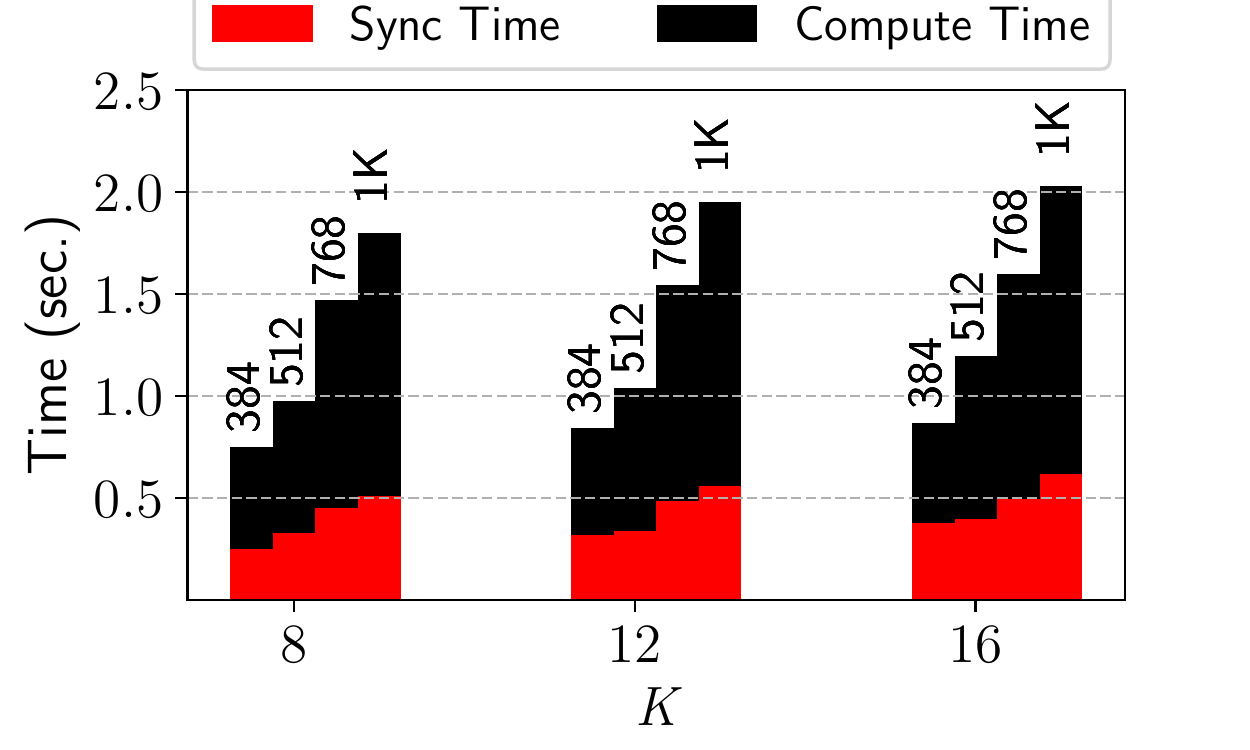}}	
  		\subfloat[ResNet50 \label{fig:resnet50-computesynctime}]
  		{\includegraphics[width=0.25\textwidth]{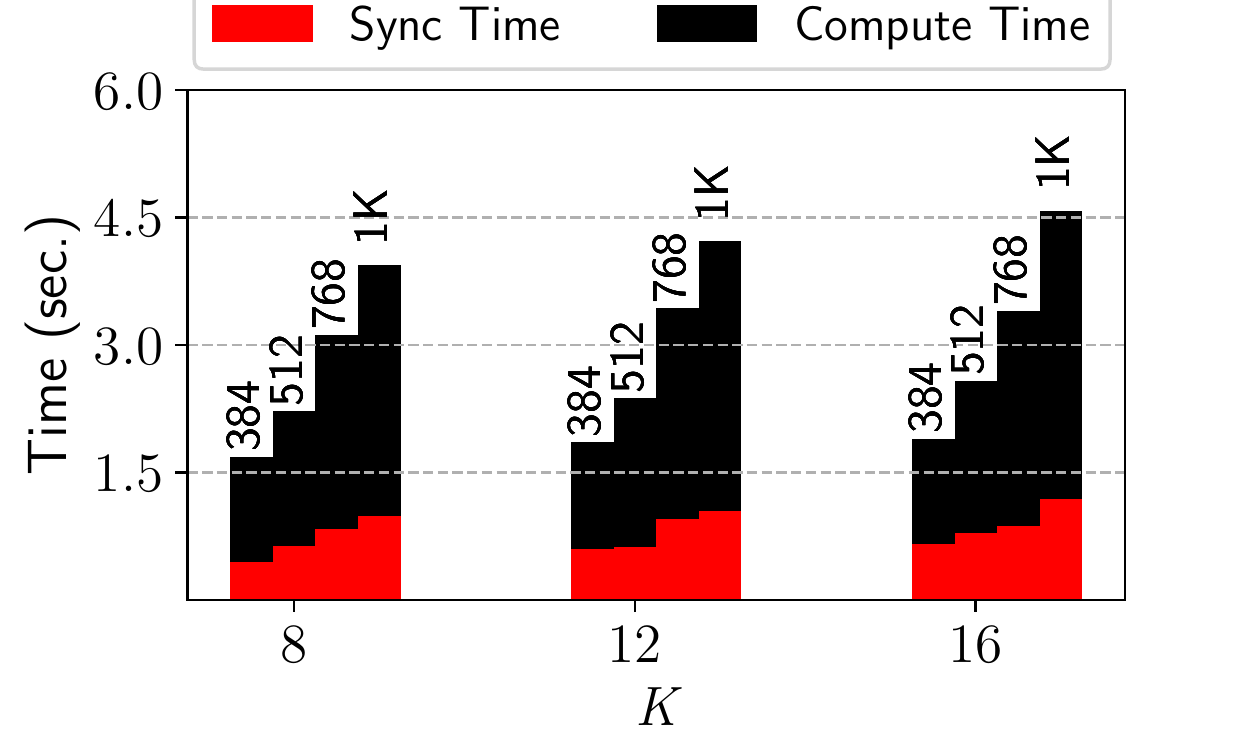}}
  		\subfloat[Transformer Base \label{fig:transformer-computesynctime}]
  		{\includegraphics[width=0.25\textwidth]{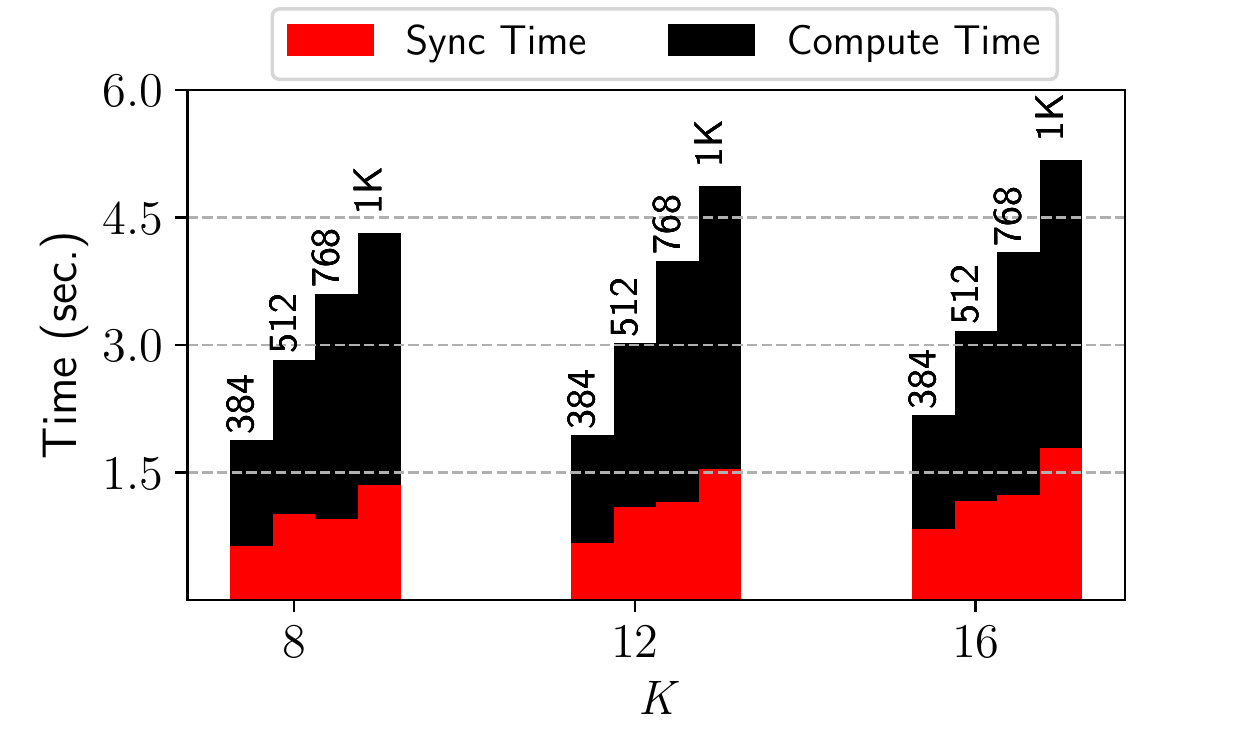}}

	\caption{The gradient computation and synchronization time breakdown for various ML workloads across multiple $K$ and trained on various $\mathit{B}$. Weak-scaling scenario: the cumulative cluster capacity is same across all $\mathit{K}$, and the worker VM size is varied.}
	\label{fig:compute-sync-time}
          \vspace*{-0.5cm}
\end{figure*}

The simplest way of scaling a parallel training job is to add more workers ($K$). 
Figure~\ref{fig:cluster-scaling} shows the decrease in the total training time to reach a fixed accuracy level for three ML models.
As the number of workers increases, the training time reduces, but there are diminishing returns.
Increasing the workers from 4 to 16 ($4\times$) only reduces the training time from 15 to 5 hours ($3\times$).

Thus, ML training shows parallel inefficiency due to the communication and synchronization overheads.
A single model-training iteration consists of a local gradient computation step, and a synchronization step where the gradients are aggregated/averaged. 
Figure~\ref{fig:compute-sync-time} shows the breakdown of this computation and synchronization.
It also shows the overhead of horizontal scaling in terms of higher synchronization overhead with increasing workers.
Here, the cumulative cluster capacity is same across various $\mathit{K}$, i.e., total CPU cores and memory allocated over all the workers in a cluster is held constant. 
We can see that increasing the number of workers increases the synchronization time. 
%For these experiments, we are using a parameter-server architecture, which minimizes synchronization overheads compared to more direct all-reduce synchronization.
With parameter servers, more workers means more stragglers, and because bulk synchronous processing is used, this increases the communication costs for everyone.
The larger number of workers also increases the work for the parameter servers, which increases the synchronization time further. 
Figure~\ref{fig:compute-sync-time} also shows the breakdown for different batch sizes. 
%Here, the cumulative cluster capacity is same across various $\mathit{K}$, i.e., total CPU cores and memory allocated over all the workers in a cluster is same regardless of the cluster size.
We can see that the gradient computation time also increases with increasing batch sizes.

\noindent \textbf{Memory.}
The VM's memory size is also an important resource for model training.
Insufficient memory forces smaller batch sizes, which reduce the training accuracy and require more iterations and synchronization during model training.
Figure~\ref{fig:memory-limit} shows the final model accuracy reached when training the Transformer model under a strict time deadline.
The smaller VMs provide insufficient accuracy, and below a 4GB threshold, the system (TensorFlow) crashes and makes no forward progress at all.

\subsection{Vertical Scaling: Increasing Batch Size}

\begin{figure}
\hspace*{-0.5cm} 
\includegraphics[width=0.52\textwidth]{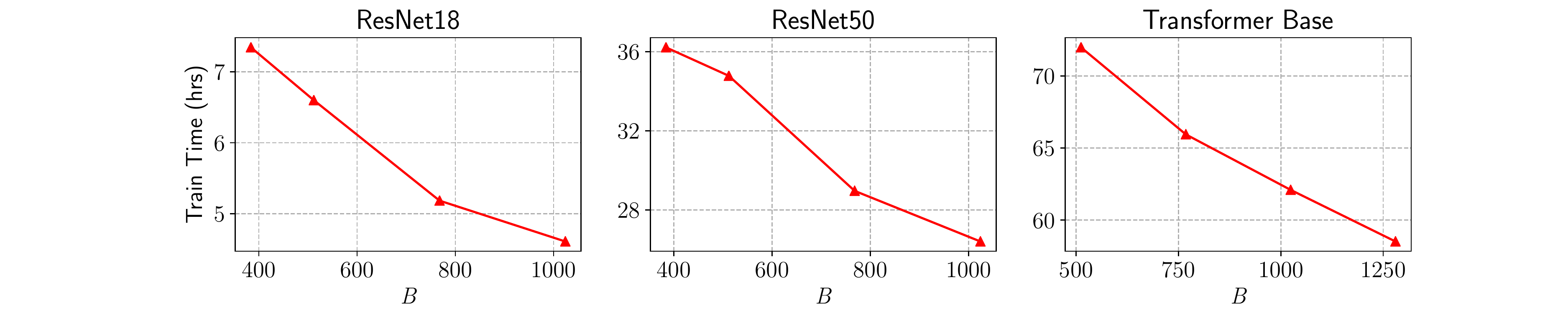}
\caption{Training time to convergence for various global batches on $\mathit{K=16}$.  ResNet18, ResNet50 and Transformer are trained to 80\%, 90\% accuracy and 18.0 BLEU score respectively.}
\label{fig:time-multipleB}
  \vspace*{-0.5cm}
\end{figure}

One way to reduce the synchronization overheads is to increase the batch size, which reduces the total number of iterations required and increases the parallel efficiency.
This is illustrated in Figure~\ref{fig:loss-steps-examples}, which shows the training-loss for different batch sizes for the ResNet18 model. 
Larger batch sizes (1024) achieve lower (i.e., better) loss compared to smaller ones. 
This is also seen for other models in Figure~\ref{fig:time-multipleB}, which shows the training time to desired accuracy for $K=16$. 
%ResNet18, ResNet50 and Transformer Base are trained to 80\%,90\% training accuracy and 18.0 BLEU score respectively. 

\begin{figure}
\hspace*{-0.45cm}
\includegraphics[width=0.52\textwidth]{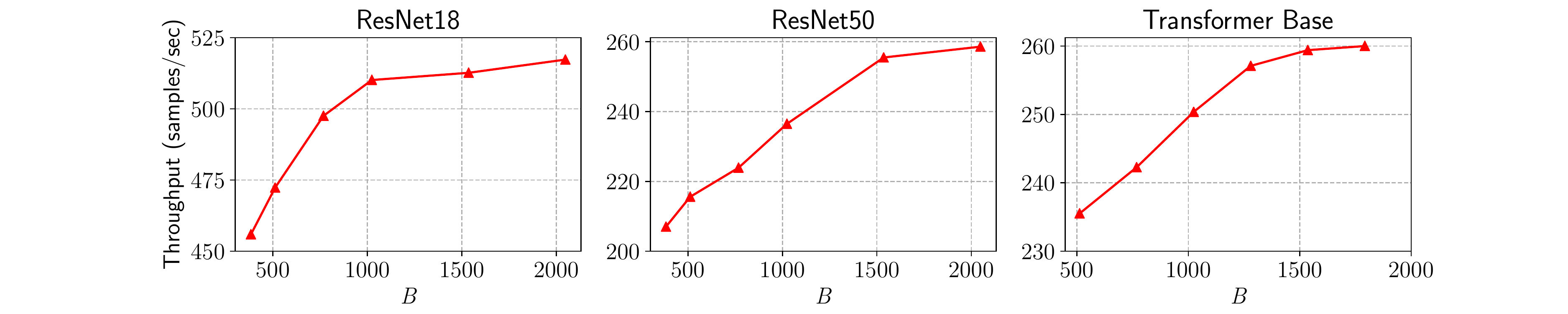}
\caption{Throughput of various models on increasing the global batch size B for $\mathit{K=16}$. The throughput increases as we increase $\mathit{B}$ upto a certain point, then plateaus.}
\label{fig:throughput-B}
  \vspace*{-0.5cm}
\end{figure}

The gains of compute efficiency with larger $B$ is evident from Fig. \ref{fig:throughput-B} where the throughput (i.e., the number of samples processed per second in the training phase) increases as the global batch size increases, then saturates after a knee-point/inflection point. 
The throughput plateaus after a certain batch size since the CPU utilization of the workers maxes out after the inflection point.
In the results shown, the workers used in the cluster are GCP E2-standard VMs with 4 vCPUs and 16 GB memory. 

%Worker with higher compute units could increase the throughput as well as the global batch size corresponding to the inflection point. Nevertheless, it is still bounded by the parallel scaling limitations of the individual workers as governed by Amdahl's law, and shown in \S\ref{subsubsec:amdahl}.

%%%%%%%%%%%%%%%%%%%%%%%%%%%%%%

\subsection{Statistical inefficiency} % and Gradient Noise}
\label{subsec:statsineff}

In both horizontal and vertical scaling, parallel training does not scale linearly. 
The fundamental reason for this non-linear scaling is that not all computing work is effective because of stochastic gradient descent. 
%Each worker computes gradients only on a subset (mini-batch) of data, and averaging diverse gradients can slow down the convergence.
In conventional parallel applications, all work performed by all workers is equally useful.
However, with stochastic gradient descent, the work done by the workers (i.e., the gradients computed) does not fully compose.
That is, the total forward progress made (i.e., the decrease in training loss) is not equal to the sum of progress made by the individual workers.
We call this the \emph{statistical inefficiency} of parallel model training, and it reflects how ``aligned'' the computed gradients of the different workers are. 
This statistical inefficiency is a fundamental attribute of SGD (and all optimization algorithms in the SGD family like Adam~\cite{kingma2014adam} etc).
The statistical inefficiency can be captured by computing the SGD noise, which is the variance in the gradients among the workers~\cite{mccandlish2018empirical,johnson2020adascale}.
We use a similar variance formulation in our model described in the next section. 

%Thus, not all work (i.e., gradient computation) is equally effective, and thus parallel training exhibits \emph{statistical inefficiency.} 

Thus adding more workers (increasing K) can increase the divergence between gradients and require more training iterations and increase the overall training time.
Similarly, a small batch size means that the gradients are computed on a small subset of data, and are more likely to differ from each other.
Thus, larger batch sizes are preferable from a statistical efficiency perspective, but have other tradeoffs: they impose additional memory requirements and communication overheads. 
Furthermore, increasing batch sizes may have diminishing returns (Figure~\ref{fig:throughput-B}).
\emph{Because of gradient noise and statistical inefficiency, throughput (number of training data samples processed per second) is not sufficient to capture the performance,} and we need to consider the wall-clock time to reach the desired accuracy level.

Our performance model is able to capture both the statistical and parallel efficiency associated with different horizontal and vertical scaling configurations, and provide accurate estimates of training time for different configurations which can be used to select the ``best'' configuration.

\section{Design}
% Modeling Parallel and Statistical Scaling}
\label{sec:design}
At  a high level, our goal is to find the ``best'' cloud cluster configuration for a model training job, with minimal information about the ML model, and in an online manner with minimal apriori profiling. 
We want to minimize the time and cost of training a model to a specified accuracy level.

% XXX 
%We decompose the cloud configuration problem into two tasks.
%Our first task is finding the optimal training \emph{job configuration} parameters, which primarily comprise of the number of parallel workers and the SGD batch size. This task is the focus of this section. 
%The second task is mapping these job configuration parameters to the cloud configuration, which consists of the VM configuration and the cluster size. We focus on this task in the next section. 

For optimizing the job configuration for a given ML training job, we first develop an analytical performance model for estimating the total training time (and cost).
This performance model is used to compute the time vs. cost tradeoff curves for a job, which can be used to select the ``right'' cloud cluster based on user preferences and constraints.
Our focus is on building simple, practical, and generalizable performance models that do not require offline training, and which can be refined and used with  online profiling. 
Predicting the total training time of ML training is especially challenging due to the statistical inefficiency of distributed SGD. 
To address this challenge, we investigate and use general \emph{SGD noise} indicators, that serve as a proxy for statistical inefficiency (Section~\ref{subsec:sgdnoise}). 
Using these scaling indicators, we develop an analytical \emph{statistical} performance model, which we combine with a more conventional parallel performance model. 
Finally in Section~\ref{subsec:alloc-policies}, we describe how the combined parallel and statistical performance model can be obtained and used in practice.

\subsection{SGD Noise as a Scaling Indicator}
\label{subsec:sgdnoise}

% The need for universal scaling indicators for distributed ML training. 
We have seen that simply adding more resources to a distributed training job doesn't decrease the training time uniformly.
This inefficiency is crucial in cloud environments, since it increases costs without proportional decrease in training time.
We seek a general ``scaling indicator'' which serves as a proxy for the overall parallel efficiency. 
For example, such a scaling indicator should indicate the scenarios in which adding more resources would not decrease training time, and we should stop scaling. 
Because we want online cluster optimization, this scaling indicator should also be easily computable at run-time, and be independent of the ML model and cluster size. 

For classic parallel applications, the communication and synchronization overheads typically serve as scaling indicators.
For example, we can compute the scaling efficiency as the fraction of time spent in communication, and stop scaling if this fraction increases above a threshold. 
Amdahl's law and other parallel scaling laws can then use these communication overheads and inform us about the performance and scaling properties of the application. 
Communication and synchronization overheads are also applicable for ML training and can be used to model their parallel efficiency. 
However, they are not sufficient, because of the statistical inefficiency of parallel ML training. 

\emph{Just as communication overheads can indicate parallel scaling in conventional parallel applications, are there similar scaling indicators for statistical inefficiency?}
We seek a \emph{general} indicator for statistical efficiency that is independent of the model and the execution environment (number and type of workers, etc.).
For example, such a scaling signal could indicate the batch size threshold for a given cluster size, beyond which scaling the application does not significantly reduce the training time. 

Fundamentally, the statistical inefficiency arises because of the noise in the gradients computed by the workers. 
Our main observation is that the SGD noise can be captured by the \emph{variance} in the gradients computed by the workers, and this serves as a useful general statistical inefficiency indicator.
This variance/noise can be computed by:
\begin{equation}
  %	\mathit{noise(\mathit{g_{\mathit{t}}})} =
\gamma(t) =   \dfrac{\mathbb{E}[\dfrac{1}{\mathit{K}}\sum_{\mathit{k}=1}^{\mathit{K}}\lvert\lvert\mathit{g^{\mathit{(k)}}_{\mathit{t}}}\lvert\lvert^2]}{\mathbb{E}[\lvert\lvert\mathit{\tilde{g}_{\mathit{t}}}\lvert\lvert^2]},
	\label{eqn:noise-eqn}
\end{equation}
where $\mathit{g_{t}^{(k)}}$ is gradient computed on worker $\mathit{k}$ at iteration $\mathit{t}$ and $\mathit{\tilde{g}_{t}}$ is the aggregated gradient norm obtained by reducing gradients on the parameter servers in a cluster with $K$  workers.
This SGD gradient variance has been investigated previously~\cite{johnson2020adascale, mccandlish2018empirical} to understand either batch size scaling or worker scaling, and we generalize it to both types of scaling. 

The noise, which is essentially the deviation in the calculated gradients from the ``true'' gradient, is also a \emph{practical} scaling indicator. 
It can be easily computed in the data-parallel parameter server strategy during the model training, i.e., in an online manner. 
The per-worker and aggregated gradients are collected from all workers and parameter servers respectively. 
Thus, from equation~\ref{eqn:noise-eqn}, we can compute the gradient noise by computing the ratio of the mean of the workers' local gradient norms and the aggregated gradient norm. 

\begin{figure}
  \centering
\hspace*{-1cm}  \includegraphics[width=0.58\textwidth]{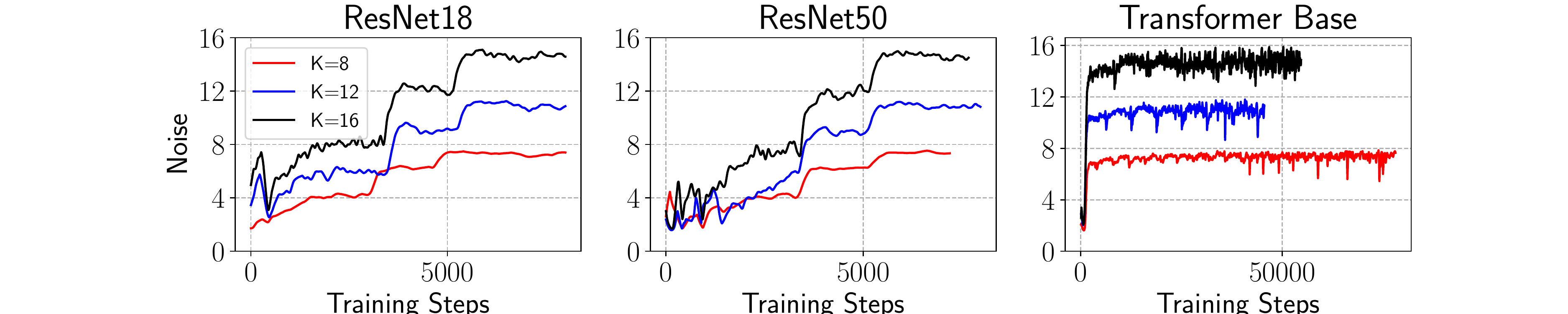}
  \caption{The SGD noise requires some iterations to stabilize, after which it is dominated by the number of workers.}
  \label{fig:noise-t}
    \vspace*{-0.5cm}
\end{figure}

In the early training stages, the variance in the gradients is on the same scale as the gradients itself and thus the initial noise is low (Figure~\ref{fig:noise-t}). 
As the ML model converges, the gradients approach towards the true gradient, increasing the noise before finally saturating to the number of workers $K$.
Since we want to compare the noise for different $K$, we normalize it by $K$, so that it is a true statistical efficiency indicator.

We have observed that the noise is not constant over the course of training, even with a static job configuration. 
Instead, the noise increases and then stabilizes, as we can see from Figure~\ref{fig:noise-t}. 
This is a fundamental artifact of SGD-based optimization, and applicable for all models and configurations.
The noise is also affected by the SGD learning rate, and we need to account for the learning-rate schedule. 
For our  cluster optimization, we want to search and select for the right cluster configuration as quickly as possible after the training commences.
However since the noise from early training epochs is unreliable,  we let the noise stabilize before using it as a scaling indicator.
When a job starts, we run it on the starting configuration until the noise stabilizes, and then begin the exploration/search process. 
This increases the overall profiling and search time, since the early iterations are the ``cold start'', but provides reliable noise estimates.

%We thus normalize the noise by $\mathit{K}$, and use this normalized noise everywhere. 

%(i.e., \emph{normalized\_noise}) to compare across different $\mathit{(K,B)}$ configurations.

\begin{figure}
  \hspace{-0.5cm}
\includegraphics[width=0.55\textwidth]{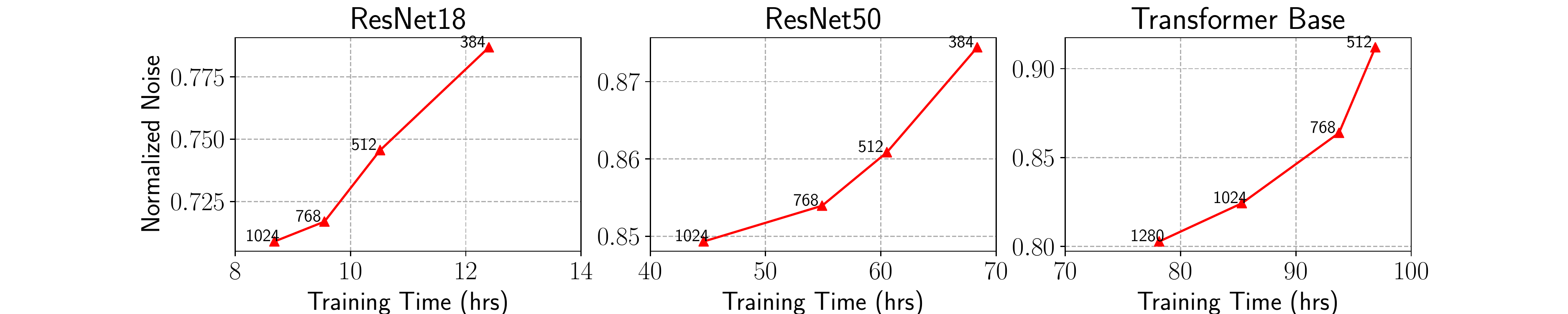}
\caption{The normalized SGD noise directly impacts the total training time for different batch sizes and models. ResNet8, ResNet50 and Transformer are trained to 80\%, 90\% accuracy and 18.0 BLEU score respectively.}
\label{fig:noise-traintime}
  \vspace*{-0.5cm}
\end{figure}
 
\noindent \emph{How effective is SGD noise in predicting performance?}
Figure~\ref{fig:noise-traintime} shows the total training time to the desired accuracy vs. the SGD noise for different global batch sizes $B$.  
For all the three ML models, the increase in noise leads to an increase in training time. 
We also observe that smaller batches have higher noise. 
Thus, the SGD noise can serve as a good indicator of the training time and efficiency.
We investigate a deeper relation of noise with statistical efficiency in the our performance model developed in the rest of this section.

\subsection{Performance and Cost Model}
\label{subsec:trainmodel}

%Our goal is to use noise in a coarse-grained way to find the scaling threshold, and not the actual correlation. 
%Additionally, we seek to do this in a black-box and online manner \emph{without} running the training on different batch and cluster sizes to completion.

We develop an analytical model for the total training time and cost of distributed ML training, which creates the tradeoff curves  (like in Figure~\ref{fig:tradeoff-example}), and guides the cloud resource allocation policies. 
Our performance models use statistical and parallel scaling indicators which can be obtained by profiling in an online manner during job execution, and do not need a-priori offline profiling.
The job's performance depends on its configuration, which consists of the number of parallel workers, $K$, and the total batch size $B$, and our model predicts the performance for each combination of these configuration parameters.  
ML training is an iterative process, and the total training time, $T$:
\begin{equation}
  % \mathit{training\_time} = \mathit{iterations}\times\mathit{itr\_time}
  T = n_i \tau,
  \label{eqn:time-model}
\end{equation} 
where $n_i$ is the number of iterations required to reach the specified model accuracy, and $\tau$ is the per-iteration time.
The number of iterations depends on the total number of training epochs $e$:
\begin{equation}
  % \mathit{iterations} = (\mathit{D}\times\mathit{epochs})/\mathit{B}
  n_i = \frac{e D}{B}, 
  \label{eqn:itr-epoch}
\end{equation}
where $D$ is the fixed dataset size, and $B$ is the global batch size, an important job configuration parameter.
The number of epochs to reach the desired model accuracy $e$, is the key unknown, and depends on many factors such as the model size, complexity, and desired accuracy, and the statistical inefficiency. 

The other key parameter in Equation~\ref{eqn:time-model} is $\tau$, which is the per-iteration time.
For a given job configuration, i.e., fixed $(K,B)$, the time to process a mini-batch is roughly constant over the course of training, because the same gradient computation and communication steps are being performed on the same mini-batch of identically distributed data.

Finally, the total cost is simply the product of training time, the number of workers $K$, and the per-VM price $p$:
\begin{equation}
  %	\mathit{cost} = \mathit{training\_time}\times\mathit{K}\times\mathit{perVMcost}
  \mathcal{C} = T K p
  \label{eqn:cost-model}
\end{equation}

We estimate the number of epochs, $e$ using our statistical performance model described in the next subsection. 
The time per iteration $\tau$, will be estimated using our parallel performance model in Section~\ref{subsec:parallel}. 

\noindent \textbf{Online Profiling and Searching.}
Using the model, we first obtain the tradeoff curves in our search or exploration phase. 
In the search phase, we  briefly run the job on some configuration, observe its parallel and statistical scaling indicators, and estimate the time (and thus cost) on that configuration.
Only a small number of iterations (around 20) are usually required for estimating the performance of a given configuration, after which we checkpoint the model, and run the job on a different configuration. 
This exploration of the various configurations allows us to obtain the full time and cost curves.

Note that due to checkpointing, there is no lost work.
The search cost is running the job briefly on suboptimal configurations, and the small overhead of restoring the model from checkpointed weights. 
Selecting the next configuration in the exploration phase is done using grid search guided by the optimization criteria and constraints on K and B. 
We refer to this as a \textbf{full or offline search},  since we first explore the configuration space, and then run the reminder of the job on the best configuration. 

To reduce the search cost, our phenomenological statistical and parallel performance model also allows us to estimate the running time on configurations without even profiling on them. 
That is, we can obtain estimates of T by profiling on only a small sample of K, B configurations, and use our phenomological models to build the rest of the tradeoff curve by fitting the learned the performance models. 
This \textbf{partial or online search} reduces the search cost significantly.
However, the drawback is that the estimates of running time due to the interpolation/regression can be error-prone, and thus the tradeoff curves obtained using the online search can differ slightly from the offline search.

Finally, we observe that many jobs train nearly identical models as part of hyperparamter tuning, neural architecture search, etc.
For example, the hyperparamter tuning may involve dozens of jobs that train the same model, but with different activation functions, weight decay, regularization, etc. 
%the learning rate schedules etc.
In such cases, the parallel and statistical efficiency of the job doesn't significantly change. 
Thus, once a job's performance model is learnt, it can be stored and \emph{reused} when the same or similar model is trained in the future.
We can thus avoid the exploratory search phase entirely, and this is the \textbf{no-search} scenario. 
We develop full, partial, and no-search techniques for both the statistical and parallel performance models.

\subsection{Statistical Performance Model}
\label{subsec:statistical}

\begin{figure*}
  \centering 
\includegraphics[width=0.75\textwidth]{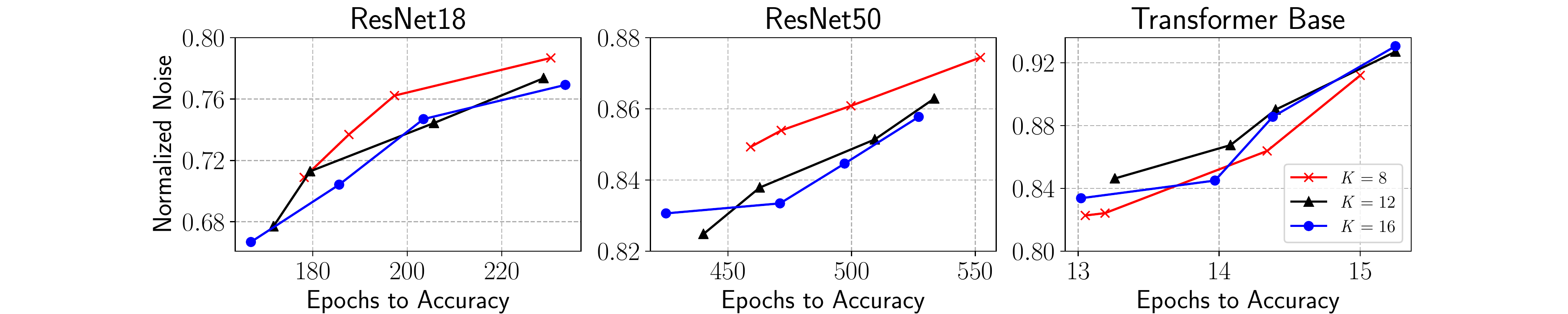}
\caption{The number of epochs required to reach various accuracy levels is linear in the normalized SGD noise.}
%  Normalized noise vs epochs for various workloads. We observe there is near-linear relation between noise and epochs.}
\label{fig:noise-epochs}
  \vspace*{-0.5cm}
\end{figure*}

The SGD noise scaling indicator allows us to model the statistical performance and the number of epochs required for achieving the desired accuracy level. 
The SGD noise increases the total amount of work required, and hence the number of iterations and epochs.  
For a given job configuration $(K, B)$, the number of epochs $e$ is proportional to the SGD noise $\gamma$  :
\begin{equation}
  e \propto \gamma ,
  \label{eqn:normnoise-epochs}
\end{equation} 
Empirical support for this can be seen in Figure~\ref{fig:noise-epochs}, which shows the normalized noise plotted against epochs taken to reach a specific performance target across various $K$ and for different target accuracy levels.

%(80\% training accuracy for ResNet18 on CIFAR10, 90\% for ResNet50 on CIFAR100 and 18.0 BLEU score for Transformer Base on WMT'14 dataset).

This linear model can be understood in relation to full gradient descent, which has no noise, and the minimal number of epochs $e^*$.  
Thus for SGD, for a given $B$, we have $e = e^* +  \theta \gamma$, where $\theta$ is the unknown linear-model parameter which relates the noise to the statistical efficiency.

%We do not know the full gradient descent performance, $e^*$, and thus cannot use the above simple relation directly.

\noindent \textbf{Full offline search.}
We profile the model on each $(K, B)$ configuration, and measure the noise $\gamma_{K, B}$.
Note that we are primarily interested in the \emph{relative} performance of various configurations.
Both  $e^*, \theta$ are properties of the model and not affected by $K,B$.
We can obtain them from prior profiling runs like those shown in Figure ~\ref{fig:noise-epochs}.
From the figure, we can see that the epochs required to reach different accuracy levels is not sensitive to the number of workers.
In fact, this is a static property of the ML model itself, and not influenced by any job configuration parameter.
Thus, if we have access to any single prior execution of the model (under any configuration, and even without profiling), then we can estimate the epochs required to reach any desired accuracy level.
In many cases however, we only need to compare the \emph{relative} performance between different configurations, for which we do not need any prior execution log, and can compare the epochs of configurations based on their observed noise.

%
%the uknown parameters $e^*, \theta$ can be obtained using online profiling of the job under various $(K, B)$ configurations, since the noise is dependent on $B$, and $e^*$ is a fundamental attribute of the model and the data-set size, and doesn't change with either $K$ or $B$.

The above full search technique already provides significant new capabilities for statistical efficiency modeling.
We refine it with two more powerful insights that reduce the search cost associated with the statistical efficiency model further.

\begin{figure*}
	\centering
	\includegraphics[width=0.8\textwidth]{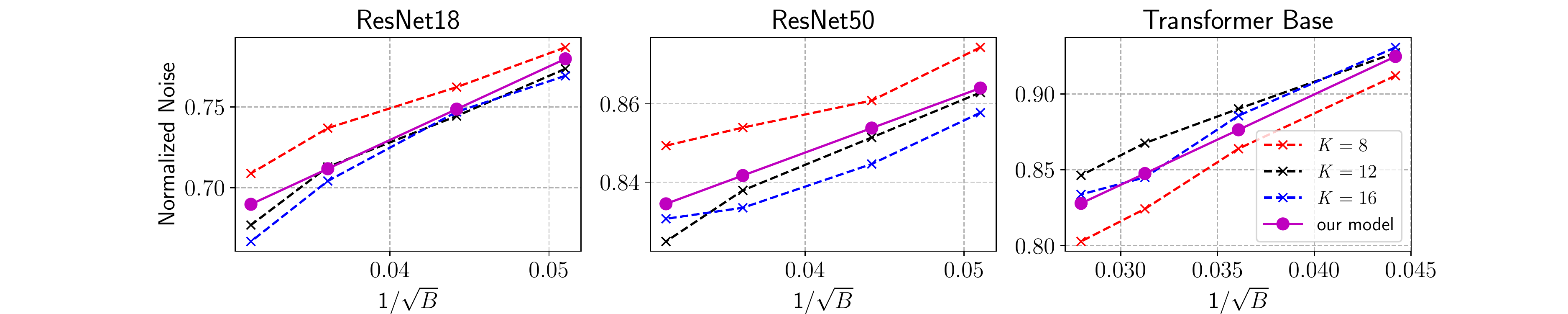}
	\caption{The noise for each $(\mathit{K},\mathit{B})$ config at 80\%, 90\%  train accuracy and 18.0 BLEU for ResNet18, ResNet50 and Transformer Base. The normalized noise is not very sensitive to $K$, and our average noise model can estimate noise for any $(\mathit{K}, \mathit{B})$ configuration with low error.}
	\label{fig:noise-B}
          \vspace*{-0.5cm}
\end{figure*}

\noindent \textbf{Partial Search.}
While the linear relation between noise and epochs is extremely powerful, we can enhance it even further to model the statistical efficiency using partial search without exploring the entire configuration space. 
First, we develop a finer-grained model for noise, which allows us to relate it to the batch size. 
\begin{equation}
  \gamma_{K,B} \propto \frac{1}{\sqrt{B}}
  %	\mathit{normalized\_noise(\mathit{g_{t}})} = \dfrac{\mathit{noise(\mathit{g}_{t})}}{\mathit{K}} = \dfrac{\alpha_{0}}{\sqrt{\mathit{B}}} + \beta_{0}
  \label{eqn:noise-rootB}
\end{equation}
This allows us to estimate the noise on different batch sizes without even requiring profiling, we can profile and find the noise on a small number of different $B$ values in the search phase, and build a linear model of $\gamma, B$, and use it to interpolate the noise for the rest of the unseen values of $B$.
This relation between $\gamma, \sqrt{B}$ is derived from the theoretical properties of SGD described in~\cite{mccandlish2018empirical}. 
We  empirically show it in Figure~\ref{fig:noise-B}, which shows the linear relation between noise and $\frac{1}{\sqrt{B}}$ for different ML models.
  
Our partial search is performed by running the job briefly on the \emph{extreme} points of B, i.e., on the smallest and largest batch size provided by the user, and then fitting a model to Equation~\ref{eqn:noise-rootB}. 

For enhanced accuracy, we can repeat the process for different values of $K$.
We have found that it is possible to avoid this per-K profiling and instead use a general/average model for noise and B. 
Surprisingly, the relation between noise and batch size is not very sensitive to the number of workers K. 
Thus, we can simplify statistical efficiency model even further, by using an \emph{average} model for noise vs. B.
This average model is also shown in Figure~\ref{fig:noise-B} (by the solid line).
For a given B, we average the (estimated) noise for various K values. 
With this averaging, using only the batch size, we can predict the noise, and thus the number of epochs.

\noindent \textbf{No-search.}
If the same or similar ML model is being trained repeatedly, then we can use its noise vs. B relation, and do not need any further profiling for modeling its statistical performance.

\subsection{Parallel Performance Model}
\label{subsec:parallel}

The above statistical performance model provides us the estimate of the number of epochs/iterations required.
We now tackle the relatively simpler task of modeling the per-iteration time, using more conventional parallel performance techniques. 
Our key insight is that ML training is highly repetitive and the performance characteristics of each iteration within a job are nearly identical.
This allows us to continue using the profiling based search strategy.
Thus the \textbf{full-search} for the parallel performance model simply runs the job for a small number of iterations on all the job configurations of interest. 

\noindent \textbf{Partial-search.}
For the partial search, we again use a phenomological model for iteration time and use an interpolation approach. 
Each iteration entails computing the gradients, collecting and averaging them, and then synchronizing them between workers via the parameter server. 
Both these major components can be modeled as follows: 
\begin{equation}
  % \mathit{iteration\_time}
  \tau = \mathit{compute\_time} + \mathit{sync\_time}.
  \label{eqn:itr-time}
\end{equation}
The gradient computation time on a worker depends on the mini-batch size, b: 
\begin{equation}
  \mathit{compute\_time} \propto \mathit{b}. 
  \label{eqn:compute-sync-time}
\end{equation}
The synchronization time is influenced by number of parallel workers:
\begin{equation}
  \mathit{sync\_time} \propto \mathit{K}
\end{equation}
Using these relations, we can build a model for the per-iteration time $\tau$ by profiling on the extreme points in the (K, B) configuration space, and then fitting linear models for the computation and synchronization. 

\noindent \textbf{No-search.} In case of repeated model training, since the computation and synchronization costs do not change, we can reuse the performance model from identical/similar models, and avoid the search phase altogether.

\subsection{Resource allocation policies}
\label{subsec:alloc-policies}

We combine the statistical and parallel performance model for our job configuration and cloud resource allocation policies.
We first build the time/cost tradeoff curves using the profiling and modeling.
Depending on the prior information available, the search strategy and costs may differ.
We have built our system as a service, so future jobs training similar models can be significantly sped-up using their stored performance models and using the partial or no-search policies.

The job configuration search is ultimately determined by the user's objective and constraints.
We support optimizing for time, cost, and also a knee-point based optimization that selects the knee-point of the cost/time curves.
We determine the knee of the curve using the kneedle~\cite{kneedle} algorithm.

Constraints on the maximum cost and time are provided by the user.
This bounds the search space and is also practical.
These constraints thus also impose a constraint on the number of worker VMs (K), and yield $K_{\text{min}}$ and $K_{\text{max}}$.
The bounds on the batch size are determined by the memory-size of the VM, yielding $B_{\text{max}}$.
Small batch sizes result in extremely high noise, and thus realistic lower-bounds on $B$ are necessary.

\section{Implementation}
\label{sec:impl}

Scavenger is implemented as a modular extension to TensorFlow, and written in Python in about 2000 lines of code.
The training scalability indicators are implemented by extending  TensorFlow's estimator framework~\cite{tf-estimator}.
Users simply need to download our TensorFlow distribution (or apply a patch), and no modifications are required to the models or any workflow component. 
The parameter server computes the SGD noise by computing the gradient norms for all the workers' updates, and the final norm for the averaged gradient.
This approximates the gradient variance, as shown in~\cite{johnson2020adascale}.
The gradient variance can be  noisy, and we use exponentially weighted moving average to smoothen the output.

All the scaling indicators: the gradient noise, gradient computation time, and synchronization time, are sent to an external model service on every iteration.
The model service uses these scaling indicators to update the performance model if operating in the initial exploratory search mode.
The user can select the full or partial search mode based on the search-cost and performance-model prediction accuracy requirements. By default, we use the partial-search, since its results are comparable to full-search with lower search costs.
Scavenger saves all performance models on persistent storage, and the no-search strategy is used if a model has been trained before. 
Once the tradeoff curves are constructed, we select the best configuration and stop all profiling. 

We interface with standard cloud APIs for managing VMs. 
Our partial-search process starts with the smallest $K,B$ configuration, and then adds more VMs to the cluster to reach the largest configuration. 
We use lightweight checkpointing: since the parameter server stores the latest model weights, the new workers in a new configuration  pull the latest weights from the parameter server and resume training. 
We switch to different configurations only on iteration boundaries, and thus no work is lost. 
The existing VMs are always reused, to avoid excessive VM churn and startup/shutdown overheads. 
Although Scavenger is currently implemented in TensorFlow v1.5, its main components are modular, and need only minimial profiling information from the ML framework. 
Supporting PyTorch is part of our ongoing work.

\section{Experimental Evaluation}\label{sec:eval}

%\noindent \textbf{Setup and workloads.} We use three commonly used ML models for training. All models are from the TensorFlow examples.
%We use VM configurations and pricing from Google Cloud.

We use popular deep learning models: two residual networks and one attention-based transformer, and evaluate across different VM size and price configurations from the Google Cloud Platform (GCP). 
Our experimental evaluation is focused on answering the following questions:
1. How effective is gradient noise as an indicator of statistical efficiency?
2. How accurate is our performance and cost model across different job configurations?
3. What are the performance and cost tradeoffs for different cloud computing cost models?
4. What are the time and cost savings achievable with our job configuration and resource allocation policies?

While most work on model training uses GPUs, we perform all evaluation on CPU VMs. 
GPUs simply reduce the per-iteration time, and all aspects of Scavenger such as the model and service are unaffected by the underlying hardware parallelism.
Standard CPU VMs can also be sized in a fine-grained manner and we can configure the VM with arbitrary amounts of CPUs and memory. 
This allows us to also evaluate \emph{weak scaling}: the total computing resources across all our cluster configurations are the same, but they are distributed among VMs differently. 
In contrast, GPUs have fixed and limited memory, and severely limit weak-scaling and  batch-size scaling. % and introduce other performance artifacts that we 
Furthermore, we only consider the worker cost, and assume that sufficient  parameter servers are launched and available. 
Parameter server allocation is tackled by other systems such as Optimus~\cite{peng_optimus_2018}, and is orthogonal and complementary to our work. % because of our focus on weak scaling. 

\begin{figure*}
  \centering 
		\subfloat[ResNet18 \label{fig:resnet18-pareto}]
  		{\includegraphics[width=0.25\textwidth]{resnet18costtime.pdf}}	
  		\subfloat[ResNet50 \label{fig:resnet50-pareto}]
  		{\includegraphics[width=0.25\textwidth]{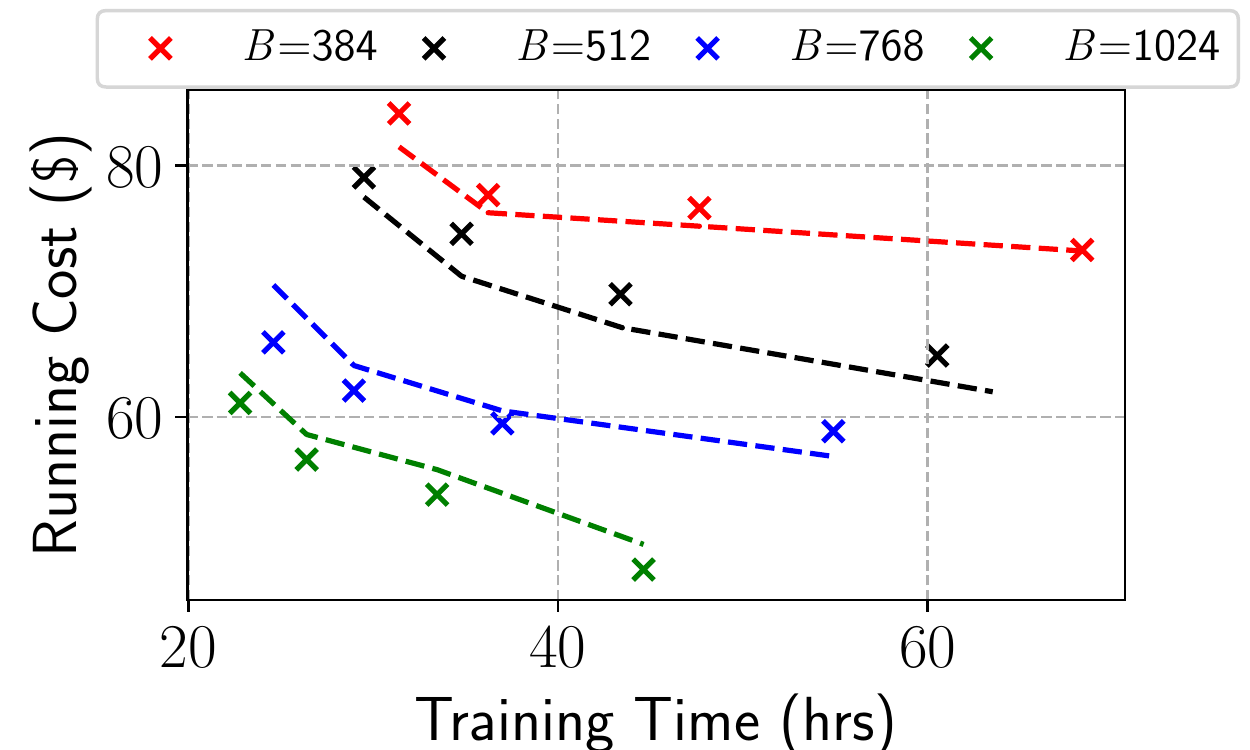}}
  		\subfloat[Transformer Base \label{fig:transformer-pareto}]
  		{\includegraphics[width=0.25\textwidth]{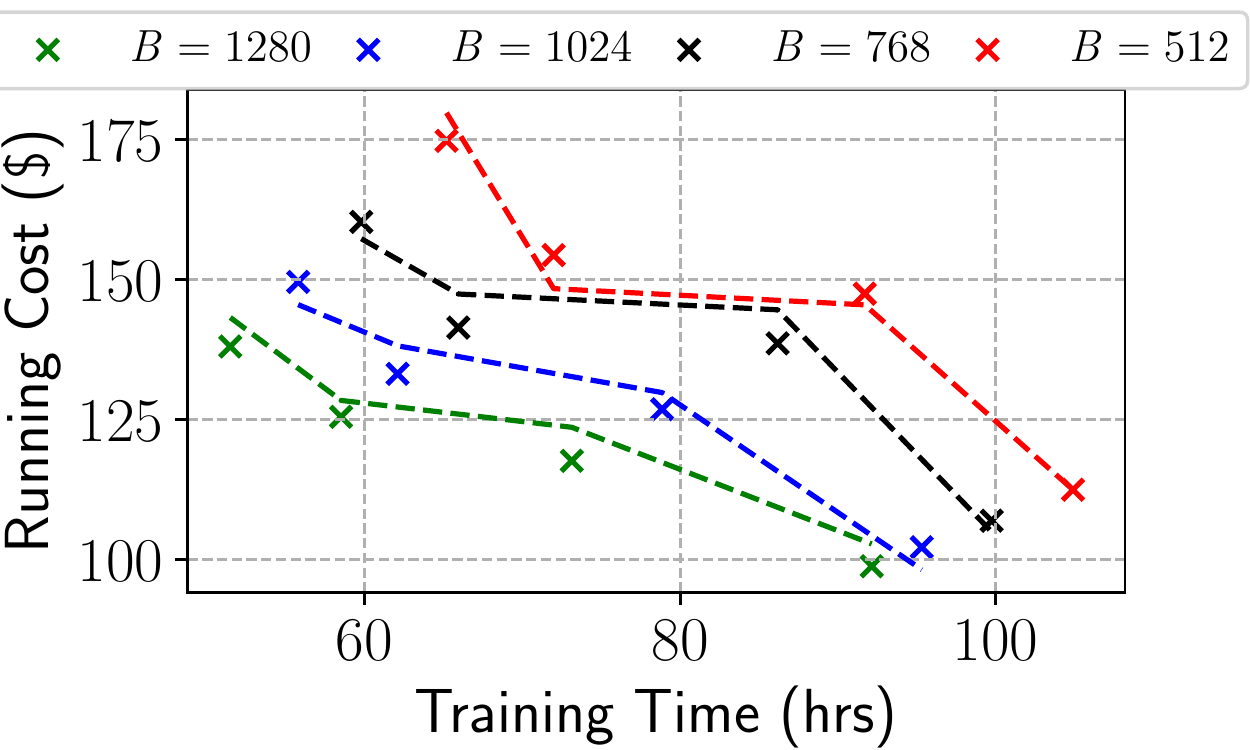}}
	\caption{Cost-Time trade-offs for ResNet18, ResNet50 and Transformer Base to reach 80\%,90\% train accuracy and 18.0 BLEU score for various $\mathit{B}$ on Google E2-standard-4 VMs.  The rental cost of each worker is fixed ($= \$0.13402/hr$).  We show the trade-offs between running cost and time for a given $B$ across decreasing cluster-sizes $[20, 16, 12, 8]$. Dashed line shows the cost predicted by our full-search performance model.}
	\label{fig:running-cost}
          \vspace*{-0.5cm}
\end{figure*}

\subsection{Cost and time tradeoffs}

With the performance model described in Section \ref{sec:design}, we can predict running cost and time for distributed training for various cluster configurations.  
Figure~\ref{fig:running-cost} shows the cost vs. time trade-offs for ResNet18, ResNet50 and Transformer Base to reach 80\%,90\% train accuracy and 18.0 BLEU score for various $\mathit{B}$ on Google E2-standard-4 VMs. 
Each scatter point show results from full runs for each $(\mathit{K,B})$ configuration and dashed line shows the predicted cost and time with the offline performance model. 
The rental cost of each worker is \$0.13402/hr.
Each point on the curve represents a decreasing cluster size, with $[20, 16, 12, 8]$ workers. 

We can see that there are clear cost vs. time tradeoffs for each batch size.
Here, the per-worker compute hardware is the same, and the per-hour total cluster-price is also proportional to $K$. 
The largest clusters have highest cost but also lowest running time.
Decreasing workers reduces cost slightly but significantly increases running time.

Both the ResNet models (Figures~\ref{fig:resnet18-pareto},~\ref{fig:resnet50-pareto}) have a single inflection/knee-point for all batch sizes, after which we see diminishing returns on cost.
\emph{For ResNet-18 $B$=384, $K$=16 represents ideal configuration since it corresponds to the knee-point.
For $B$=512, inflection point corresponds to $K$=12 so that is the ideal configuration at this batcgh-size.}
With Transformer model in Figure~\ref{fig:transformer-pareto}, we observe two inflection points corresponding to clusters $12$ and $16$ for any $B$.
We observed a notable decrease in iteration time from $\mathit{K}$ $12$ to $16$ since for the same per-worker compute hardware and $B$, since larger $K$ implies smaller worker mini-batch size.
For example, $B=768$ changes mini-batch size from 64 to 48 when $K$ goes from 12 to 16. 
Thus, we see a significant training time difference between $K=12$ vs. $16$, resulting in two distinct inflection points.  

\noindent \textbf{Result:} \emph{The tradeoff curves can be a crucial tool for judicious resource allocation on the cloud for distributed training.}

The dashed lines in Figure~\ref{fig:running-cost} shows the cost predicted by our performance model using the \emph{full-search} strategy, which relies on profiling of the gradient noise and iteration-time performance models by running the model on different configurations for a small number of iterations. 
Compared to the actual job running time, our offline performance model has an error of only 1-5\%, across the entire range of models, workers, and batch sizes.

%The memory allocated to a VM is another factor that determines the convergence of a given model, as can be seen from Fig. \ref{fig:memory-limit}.  We assumed the cost of each VM to be the same for the results presented in Fig. \ref{fig:running-cost} and eqn. (\ref{eqn:cost-model}).

\subsection{Partial Search}
We now evaluate the effectiveness of our partial search statistical and parallel performance model.
In the partial search strategy, we only profile the job on a small number of configurations (and only for a few iterations).
We then use the phenomenological models and linear regression for estimating the job running time for the other configurations. 

For our evaluation, we set $8\leq K \leq20$ and $384 \leq B \leq 1024$. 
In case of Transformers, we set $\mathit{B}_{min}$, $\mathit{B}_{max}$ to $512$ and $1280$. 
%$\mathit{B}_{min}$ and $\mathit{B}_{max}$ as $8,20,384$ and $1K$ (in case of Transformers, we set $\mathit{B}_{min}$, $\mathit{B}_{max}$ to $512$ and $1280$).
We increment $K$ by $4$ to compare results with offline runs from Figure~\ref{fig:running-cost}, so we use $K \in [8,12,16,20]$. 

The starting configurations are  $(\mathit{K}_{min},\mathit{B}_{min})$ and $(\mathit{K}_{min},\mathit{B}_{max})$, until the gradient noise has stabilized. 
With exponential moving average smoothing,  noise for ResNet18, ResNet50 and Transformer Base stabilized at $2K, 3K$ and $10K$ iterations respectively. 
The total search cost for ResNet 18, arising from doing this profiling on extreme configurations was minimal.
\emph{The overhead of exploring a new configuration (due to checkpoint-restore) is minimal, on average $37$ seconds for ResNet18,  $40$ for ResNet50, and $127$ seconds for Transformers.
Each configuration is run for around 20 iterations, which takes around 17--35 seconds for our three models. }
Compared to an ``oracle'' scenario of  running on the optimal configuration all along (bypassing the search phase), our approach increases running time by $0.83$ hours, and \$0.89 to the final cost.
This represents a 13\% increase in running time and 9\% increase in cost, compared to an oracle approach which runs the job on the optimal configuration from the start.
Compared to arbitrary job configuration without our techniques, our running times can be more than $2\times$ lower and costs can be more than 40\% lower. 

\noindent \textbf{Result:} \emph{The partial search increases job running time by 13\% and cost by 9\%, even compared to an oracle approach, and is a low-overhead strategy for discovering optimal configurations.} 

\begin{figure*}
%	\begin{subfigure}{0.9\textwidth}
          \hspace{0.2cm}
          \centering 
		{\includegraphics[width=0.35\textwidth]{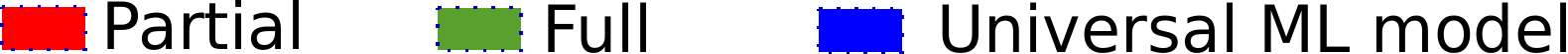}} %{errlabel.png}}
%	\end{subfigure}
		\hspace{-0.1cm}		
		\subfloat[ResNet18 \label{fig:resnet18-err}]
  		{\includegraphics[width=0.25\textwidth]{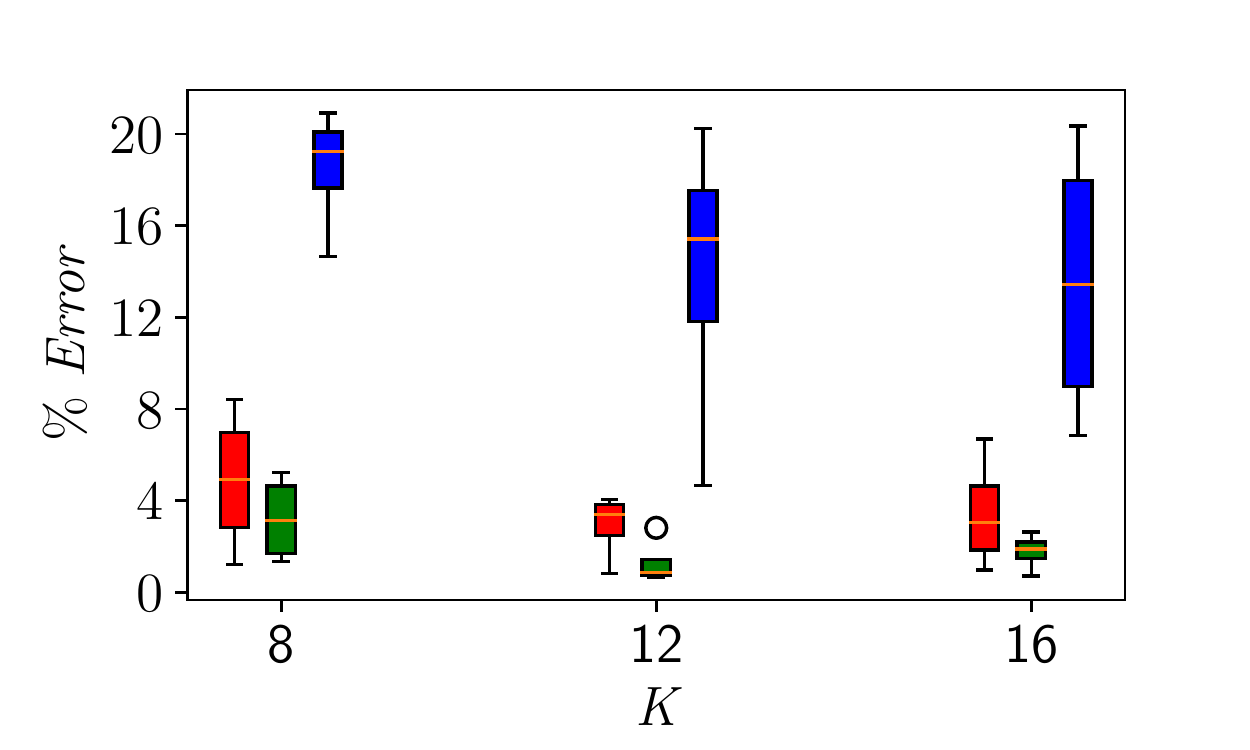}}	
  		\subfloat[ResNet50 \label{fig:resnet50-err}]
  		{\includegraphics[width=0.25\textwidth]{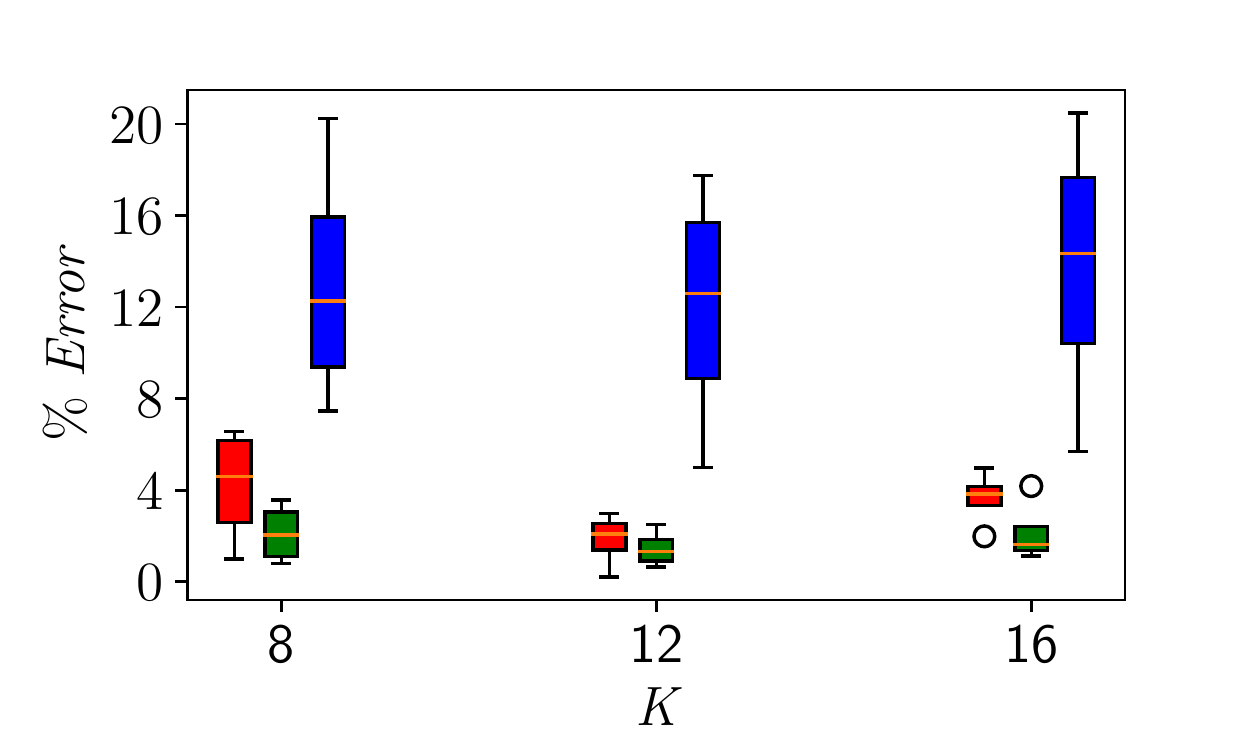}}
  		\subfloat[Transformer Base \label{fig:transformer-err}]
  		{\includegraphics[width=0.25\textwidth]{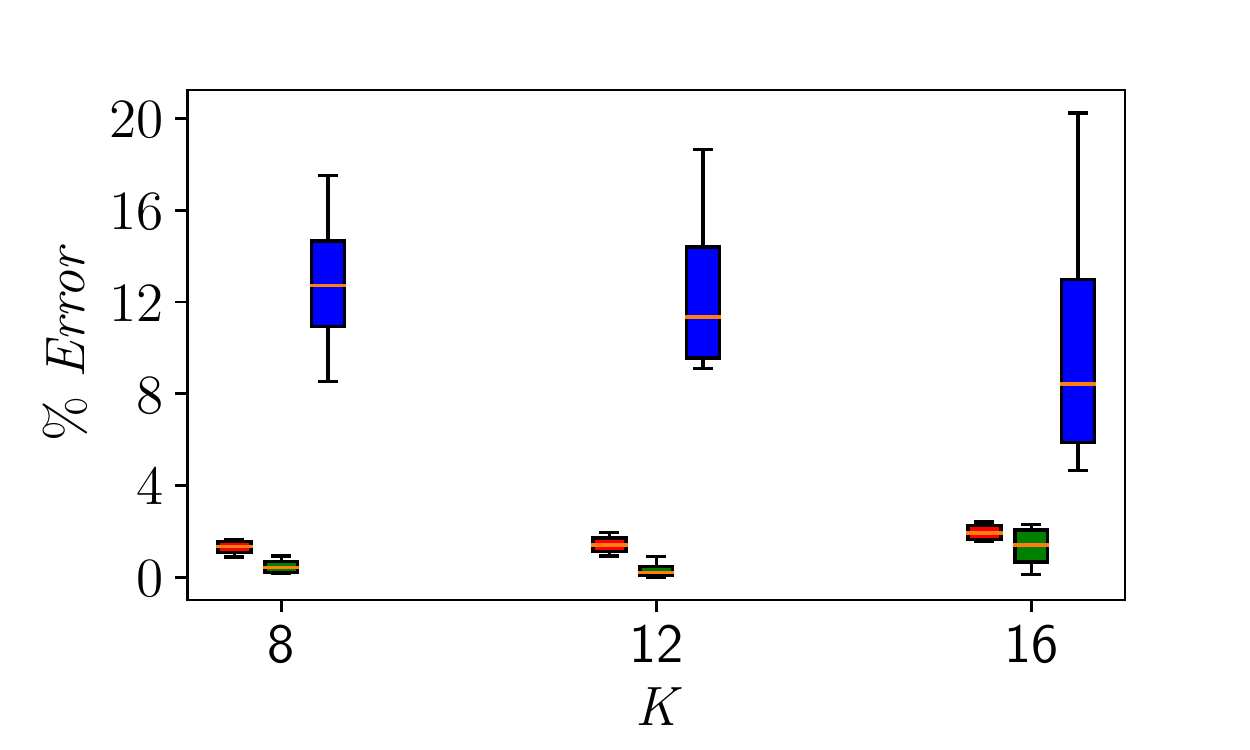}}

                \caption{Error in predicted training time from actual training time across all job configurations. The error for partial and full search is low. Even the universal model, which doesn't consider any model-specific details, provides acceptable results.}

%                  taken for various $\mathit{B}$ for each $\mathit{K}$ with offline and online approaches. Online approach has higher prediction error, but at an overall lower cost as compared to offline techniques.}
                \label{fig:on-off-err}
                  \vspace*{-0.5cm}
\end{figure*}

\subsection{Model Accuracy}

Both the full  and partial  search are able to accurately predict the total training time, as seen from Figure~\ref{fig:on-off-err}.
We evaluate three configurations: partial search (red), full search (green), and a worst-case no-search strategy.
The figure shows the distribution of the error of running time prediction vs. the empirical job running time, across different K and B.
We see that the average error for partial search is 4\% for ResNet and less than 2\% for Transformer.
The full-search is even better:  with an average error of 0.5--3.5\% across all models and configurations. 
%Note that the partial-search cost is about $4\times$ lower than the full-search cost, which is already insignificant and less than 10\% of the optimal cost. 

In  Figure~\ref{fig:on-off-err},  we also evaluate our no-search strategy in a worst-case scenario. 
The no-search strategy performance is exactly the same as the full-search scenario, if a near-identical ML model has been trained before.
However we construct a scenario where a ``global average''  performance model is used which averages the statistical and performance models \emph{over all the three ML models.} 
Thus we are using a ``universal'' performance model.
Even this universal global-average model shows acceptable training-time prediction: the error is in the range of 4--20\%. 
Note that this global-average model does not require any search, has no search costs, nor does it require any prior profiling or pilot runs. 
It is thus fully online and zero overhead.

Finally, we note that the running-time prediction error is not highly significant to our overall objective of discovering optimal configurations.
We primarily care about the \emph{relative} running times, because we only compare configurations and run the job on the best-predicted configuration.
It is likely that the best-predicted configuration remains the same even with the higher error, or the sub-optimal configuration chosen due to the errors is very close to the optimal configuration in the trade-off curve. 

\noindent \textbf{Result:} \emph{Our performance model can predict training times with a low error of  0.5--3.5\%, and only 4\% even with partial searching. In the fully-online setting, the error range is 4--20\%.}

%We explore different $(K,B)$ configurations with \emph{kspace} and \emph{bspace} parameters.  We obtain a pareto front for one $bspace$ and various $kspace$ values.  For different $bspace$, we get different pareto frontiers. The online algorithm finds the kneepoints of all pareto fronts and returns a $(K,B)$ configuration that minimizes the overall cost and time.  \emph{It is to be noted that the selected kneepoints are already optimal from cost-time trade-off perspective}.

%For ResNet18,  the kneepoint $\mathit{(K,B)}$ configurations obtained with online Scavenger scaling for various $B$ are $(12,1K),(16,768),(12,512),(16,384)$. The ideal config selected by Scavenger that minimizes $(cost\times time)$ is $\mathbf{(12,1024)}$.
%Compared to a naively chosen random config,  Scavenger reduces time by $2.17\times$ and saves cost by $1.65\times$ against config $(8,384)$, or increases time by $1.29\times$ while still decreasing cost by $1.35\times$ against a config like $(20,1024)$. Looking at the empirical runs from Fig. \ref{fig:running-cost}, we can see that the online model does indeed select the most optimal config.

%For ResNet50, Scavenger identified the config $\mathbf{(16,768)}$ with minimum $(cost\times time)$ among the inflection points $[(16,1024),(12,768),(12,512),(16,384)]$. For Transformer,  Scavenger determined the ideal config as $\mathbf{(12,1280)}$. For all ML workloads, we can from the pareto fronts in Fig. \ref{fig:running-cost} that online Scavenger is successfully able to predict the inflection points that minimize cost-time.

\begin{figure*}
  \centering 
		\subfloat[ResNet18 \label{fig:resnet18-mem-pareto}]
  		{\includegraphics[width=0.25\textwidth]{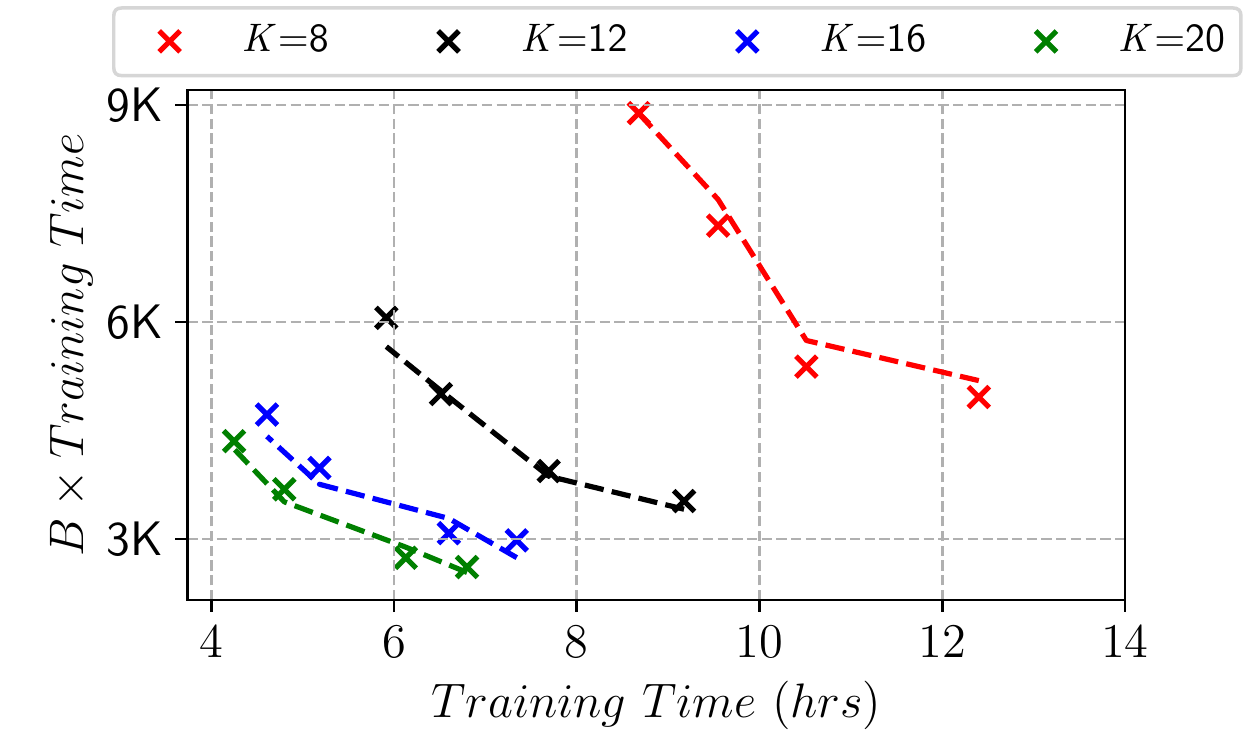}}	
  		\subfloat[ResNet50 \label{fig:resnet50-mem-pareto}]
  		{\includegraphics[width=0.25\textwidth]{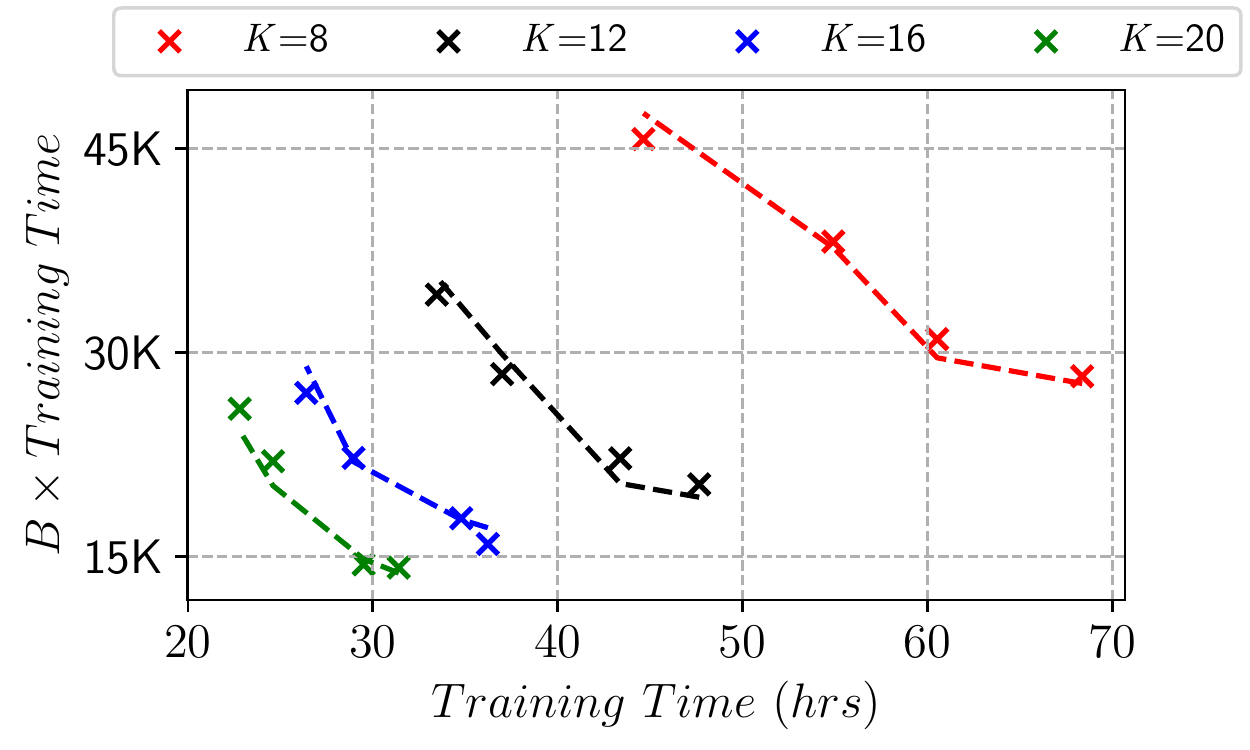}}
  		\subfloat[Transformer Base \label{fig:transformer-mem-pareto}]
  		{\includegraphics[width=0.25\textwidth]{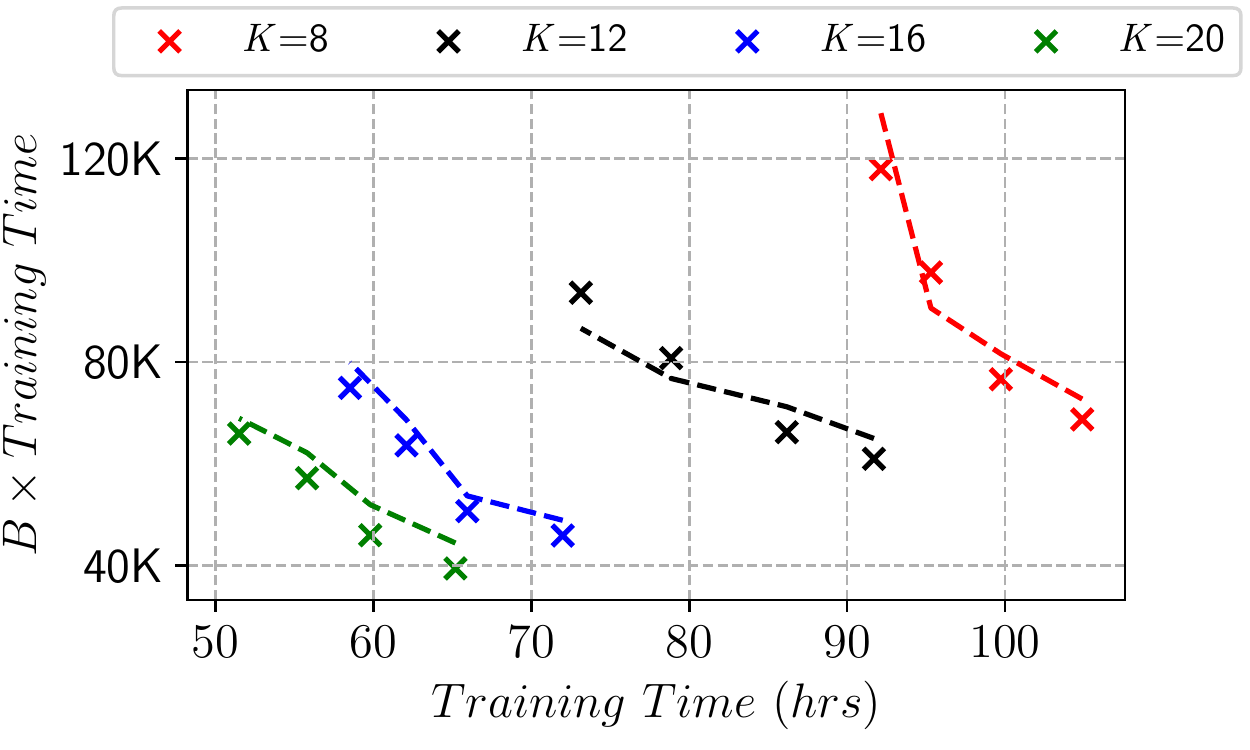}}
                \caption{Scavenger can work with different VM pricing models. In this figure, VMs are priced based on their memory size, resulting in different cost/time tradeoffs.}
%	\caption{Cost-Time trade-offs for ResNet18, ResNet50 and Transformer Base when VM prices scale linearly with memory requirements for a given $\mathit{B}$ in a model. Thus, the cost function used is $(B \times Training\ Time)$. The cost-time trade-offs are shown for a $K$ on batch sizes $[384,512,768,1024]$ for ResNet18 and ResNet50, and $[512,768,1024,1280]$ for Transformer Base. Dashed line shows the memory-bound cost predicted by the performance model.}
                \label{fig:memory-cost}
                  \vspace*{-0.5cm}
\end{figure*}

%\textbf{Accuracy of online vs. offline model:} We compare the errors in predicting training time for the configs we fully trained to 80\%,90\% train accuracy and 18.0 BLEU score for ResNet18, ResNet50 and Transformer Base. The prediction errors shown in Fig. \ref{fig:on-off-err} use eqn. (\ref{eqn:normnoise-epochs}) and (\ref{eqn:epoch-b}) for offline model,  and eqn. (\ref{eqn:normnoise-epochs}) in Algo. \ref{alg:scaling-algo} for online model. Online model predicts training time in error bounds of $\mathbf{ 4-20\%}$ for ResNet18, $\mathbf{5-21\%}$ for ResNet50 and $\mathbf{5-20\%}$ for Transformer Base. Please note that offline models were built with data from full training runs on all $\mathit{(K,B)}$ configs while the online model was partially run on a subset of configs mentioned in Algo. \ref{alg:scaling-algo}, thus giving higher errors than offline models.

\subsection{Memory-based pricing}

The cost of training is ultimately determined by the VM cost model.
So far, we have looked at conventional on-demand VM pricing, where the VM cost scales linearly according to the number of vCPUs.
Scavenger can work with different cost models.
We consider VMs that are priced both per CPU and also per GB of memory.
Google cloud's custom-sized VMs approximate this model.

With such a finer-grained cost model, the cost-time tradeoff curves are shown in Figure~\ref{fig:memory-cost}.
In this case, the VM memory is allocated according to the batch size such that there is negligible free memory.
The cost is proportional to the total memory required (the global batch size $B$) and  running time.

Comparing the results in Fig. \ref{fig:running-cost} and \ref{fig:memory-cost}, we see a shift in the inflection points for all ML workloads. This is expected since the running costs changed as a cluster $K$ training on $B=1024$ will be pricier than that running $B=384$, as more memory would be allocated to the former.

\section{Related Work}

% Most similar. Pollux kungfu
Our work falls in the category of adapting model training on distributed infrastructure such as shared clusters and cloud platforms~\cite{mayer2020scalable}.
Scavenger uses the noise scale proposed in AdaScale SGD~\cite{johnson2020adascale}, which is similar to the gradient noise model of McCandlish et.al~\cite{mccandlish2018empirical}. 
KungFu~\cite{mai_kungfu_2020} and Pollux~\cite{qiao2021pollux} also use this gradient noise metric for monitoring training performance and dynamically adjusting the resource allocation to minimize it.
In addition to elasticity and adaptation mechanisms proposed in these papers, we use a performance profiling based approach that also takes cost into account. 
KungFu is complementary to our work: we can implement Scavenger's policies as part of their adaptation-policy framework and mechanisms. 
Pollux also considers statistical efficiency and similar worker and batch-size tradeoffs, but is not cloud cost aware, and instead provides scheduling policies for shared clusters. 
BFTrainer \cite{bftrainer} attempts to utilize idle nodes for distributed training dynamically using a mixed integer linear programming (MILP) resource allocation algorithm.

Commercial offerings of ``model training as a service'', such as Amazon AWS SageMaker~\cite{liberty2020elastic}, use only rudimentary performance models, and do not use statistical efficiency or pareto-optimal allocation.
Searching for hyperparamters is an important cloud workload, and reducing this search cost using parallel search techniques and early stopping provide significant cost and time savings~\cite{dunlap2021elastic, liaw_hypersched_2019, misra2021rubberband}.
Unlike hyperparamter optimization which focuses on reducing the cost of a ``bag'' of jobs, Scavenger focuses on optimizing the cost and time of a \emph{single} job.
Efficient elasticity mechanisms and policies for ML training~\cite{li2022easyscale, qiao_litz_atc2018, narayanamurthy2013towards} can also be incorporated into Scavenger.

Scheduling and resource allocation in shared clusters is challenging for distributed training because of the complex performance tradeoffs we have identified, and the large computing requirements.
In shared private clusters, optimizing the use of limited GPU resources is a key challenge~\cite{li_ease.ml:_2017, gu2019tiresias, le2020allox, mohan2022looking}.
In cloud platforms, resource contention is not an issue, but instead cost optimization is important.

Modeling distributed ML training poses many challenges because of the heterogeneity of ML models and their performance tradeoffs~\cite{yu_computation_2019}.
Optimus~\cite{peng_optimus_2018} models the throughput and communication costs to allocate workers and parameter servers to jobs on a shared kubernetes cluster. 
Cynthia~\cite{zheng_cynthia:_2019} minimizes cloud cost and time by scaling workers and parameter servers using a finer-grained analytical model, but does not consider batch sizes and statistical efficiency. 
We do not adjust the number of parameter servers and assume that they are suitably provisioned.
Optimizing parameter server allocation is part of our future work.  
Batch-size adaptation can be important for model generalizability and performance, and can benefit from second-order gradient information~\cite{yao_large_2020}.

\section{Conclusion}
The training time and cost for large machine learning models is significant, and sensitive to many job and cloud configuration parameters. 
Scavenger is a cloud service which uses online profiling and new performance models for estimating the training performance on different cloud configurations, with high accuracy of over 95\%, and reduces training time by $2\times$. 

\bibliographystyle{IEEEtran}
\bibliography{ml}

\begin{thebibliography}{10}
\providecommand{\url}[1]{#1}
\csname url@samestyle\endcsname
\providecommand{\newblock}{\relax}
\providecommand{\bibinfo}[2]{#2}
\providecommand{\BIBentrySTDinterwordspacing}{\spaceskip=0pt\relax}
\providecommand{\BIBentryALTinterwordstretchfactor}{4}
\providecommand{\BIBentryALTinterwordspacing}{\spaceskip=\fontdimen2\font plus
\BIBentryALTinterwordstretchfactor\fontdimen3\font minus
  \fontdimen4\font\relax}
\providecommand{\BIBforeignlanguage}[2]{{%
\expandafter\ifx\csname l@#1\endcsname\relax
\typeout{** WARNING: IEEEtran.bst: No hyphenation pattern has been}%
\typeout{** loaded for the language `#1'. Using the pattern for}%
\typeout{** the default language instead.}%
\else
\language=\csname l@#1\endcsname
\fi
#2}}
\providecommand{\BIBdecl}{\relax}
\BIBdecl

\bibitem{alipourfard2017cherrypick}
O.~Alipourfard, H.~H. Liu, J.~Chen, S.~Venkataraman, M.~Yu, and M.~Zhang,
  ``$\{$CherryPick$\}$: Adaptively unearthing the best cloud configurations for
  big data analytics,'' in \emph{14th USENIX Symposium on Networked Systems
  Design and Implementation (NSDI 17)}, 2017, pp. 469--482.

\bibitem{johnson2020adascale}
T.~B. Johnson, P.~Agrawal, H.~Gu, and C.~Guestrin, ``Adascale sgd: A
  user-friendly algorithm for distributed training,'' 2020.

\bibitem{mccandlish2018empirical}
S.~McCandlish, J.~Kaplan, D.~Amodei, and O.~D. Team, ``An empirical model of
  large-batch training,'' \emph{arXiv 1812.06162}, 2018.

\bibitem{mai_kungfu_2020}
L.~Mai, G.~Li, M.~Wagenländer, K.~Fertakis, A.-O. Brabete, and P.~Pietzuch,
  ``\BIBforeignlanguage{en}{{KungFu}: {Making} {Training} in {Distributed}
  {Machine} {Learning} {Adaptive}},'' \emph{\BIBforeignlanguage{en}{USENIX
  Symposium on Operating Systems Design and Implementation}}, p.~19, 2020.

\bibitem{qiao2021pollux}
A.~Qiao, S.~K. Choe, S.~J. Subramanya, W.~Neiswanger, Q.~Ho, H.~Zhang, G.~R.
  Ganger, and E.~P. Xing, ``Pollux: Co-adaptive cluster scheduling for
  goodput-optimized deep learning,'' \emph{15th $\{$USENIX$\}$ Symposium on
  Operating Systems Design and Implementation ($\{$OSDI$\}$ 21)}, 2021.

\bibitem{sgd}
L.~Bottou, ``Large-scale machine learning with stochastic gradient descent,''
  in \emph{COMPSTAT'2010}, pp. 177--186.

\bibitem{ruder2017overview}
S.~Ruder, ``An overview of gradient descent optimization algorithms,''
  \emph{Arxiv 1609.04747}, 2017.

\bibitem{torsten_demystifying_2018}
T.~Ben-Nun and T.~Hoefler, ``\BIBforeignlanguage{en}{Demystifying {Parallel}
  and {Distributed} {Deep} {Learning}: {An} {In}-{Depth} {Concurrency}
  {Analysis}},'' \emph{\BIBforeignlanguage{en}{arXiv:1802.09941 [cs]}}.

\bibitem{mayer_scalable_2019}
R.~Mayer and H.-A. Jacobsen, ``\BIBforeignlanguage{en}{Scalable {Deep}
  {Learning} on {Distributed} {Infrastructures}: {Challenges}, {Techniques} and
  {Tools}},'' 2019, arXiv: 1903.11314.

\bibitem{dean2012large}
J.~Dean, G.~Corrado, R.~Monga, K.~Chen, M.~Devin, M.~Mao, M.~Ranzato,
  A.~Senior, P.~Tucker, K.~Yang \emph{et~al.}, ``Large scale distributed deep
  networks,'' in \emph{{NIPS} '12}, pp. 1223--1231.

\bibitem{li_scaling_2014}
M.~Li, ``\BIBforeignlanguage{en}{Scaling {Distributed} {Machine} {Learning}
  with the {Parameter} {Server}},'' in \emph{\BIBforeignlanguage{en}{{ACM}
  {BigDataScience} '14}}, pp. 1--1.

\bibitem{szegedy2017inception}
C.~Szegedy, S.~Ioffe, V.~Vanhoucke, and A.~A. Alemi, ``Inception-v4,
  inception-resnet and the impact of residual connections on learning,'' in
  \emph{31st AAAI '17}.

\bibitem{he2015deep}
K.~He, X.~Zhang, S.~Ren, and J.~Sun, ``Deep residual learning for image
  recognition,'' 2015.

\bibitem{vaswani2017attention}
A.~Vaswani, N.~Shazeer, N.~Parmar, J.~Uszkoreit, L.~Jones, A.~N. Gomez,
  L.~Kaiser, and I.~Polosukhin, ``Attention is all you need,'' 2017.

\bibitem{gupta_model_2016}
S.~Gupta, W.~Zhang, and F.~Wang, ``\BIBforeignlanguage{en}{Model {Accuracy} and
  {Runtime} {Tradeoff} in {Distributed} {Deep} {Learning}:{A} {Systematic}
  {Study}},'' \emph{\BIBforeignlanguage{en}{arXiv:1509.04210}}, 2016.

\bibitem{kingma2014adam}
D.~P. Kingma and J.~Ba, ``Adam: A method for stochastic optimization,''
  \emph{ICLR}, 2015.

\bibitem{kneedle}
V.~Satopaa, J.~Albrecht, D.~Irwin, and B.~Raghavan, ``Finding a "kneedle" in a
  haystack: Detecting knee points in system behavior,'' in \emph{2011 31st
  International Conference on Distributed Computing Systems Workshops}, 2011,
  pp. 166--171.

\bibitem{tf-estimator}
H.-T. Cheng, Z.~Haque, L.~Hong, M.~Ispir, C.~Mewald, I.~Polosukhin, G.~Roumpos,
  D.~Sculley, J.~Smith, D.~Soergel \emph{et~al.}, ``Tensorflow estimators:
  Managing simplicity vs. flexibility in high-level machine learning
  frameworks,'' in \emph{ACM SIGKDD 2017}, pp. 1763--1771.

\bibitem{peng_optimus_2018}
\BIBentryALTinterwordspacing
Y.~Peng, Y.~Bao, Y.~Chen, C.~Wu, and C.~Guo, ``\BIBforeignlanguage{en}{Optimus:
  an efficient dynamic resource scheduler for deep learning clusters},'' in
  \emph{\BIBforeignlanguage{en}{Proceedings of the {Thirteenth} {EuroSys}
  {Conference} on - {EuroSys} '18}}.\hskip 1em plus 0.5em minus 0.4em\relax
  Porto, Portugal: ACM Press, 2018, pp. 1--14. [Online]. Available:
  \url{http://dl.acm.org/citation.cfm?doid=3190508.3190517}
\BIBentrySTDinterwordspacing

\bibitem{mayer2020scalable}
R.~Mayer and H.-A. Jacobsen, ``Scalable deep learning on distributed
  infrastructures: Challenges, techniques, and tools,'' \emph{ACM Computing
  Surveys (CSUR)}, vol.~53, no.~1, pp. 1--37, 2020.

\bibitem{bftrainer}
\BIBentryALTinterwordspacing
Z.~Liu, R.~Kettimuthu, M.~E. Papka, and I.~Foster, ``Bftrainer: Low-cost
  training of neural networks on unfillable supercomputer nodes,'' 2021.
  [Online]. Available: \url{https://arxiv.org/abs/2106.12091}
\BIBentrySTDinterwordspacing

\bibitem{liberty2020elastic}
E.~Liberty, Z.~Karnin, B.~Xiang, L.~Rouesnel, B.~Coskun, R.~Nallapati,
  J.~Delgado, A.~Sadoughi, Y.~Astashonok, P.~Das \emph{et~al.}, ``Elastic
  machine learning algorithms in amazon sagemaker,'' in \emph{Proceedings of
  the 2020 ACM SIGMOD International Conference on Management of Data}, 2020,
  pp. 731--737.

\bibitem{dunlap2021elastic}
L.~Dunlap, K.~Kandasamy, U.~Misra, R.~Liaw, M.~Jordan, I.~Stoica, and J.~E.
  Gonzalez, ``Elastic hyperparameter tuning on the cloud,'' in
  \emph{Proceedings of the ACM Symposium on Cloud Computing}, 2021, pp. 33--46.

\bibitem{liaw_hypersched_2019}
\BIBentryALTinterwordspacing
R.~Liaw, R.~Bhardwaj, L.~Dunlap, Y.~Zou, J.~E. Gonzalez, I.~Stoica, and
  A.~Tumanov, ``\BIBforeignlanguage{en}{{HyperSched}: {Dynamic} {Resource}
  {Reallocation} for {Model} {Development} on a {Deadline}},'' in
  \emph{\BIBforeignlanguage{en}{Proceedings of the {ACM} {Symposium} on {Cloud}
  {Computing}}}.\hskip 1em plus 0.5em minus 0.4em\relax Santa Cruz CA USA: ACM,
  Nov. 2019, pp. 61--73. [Online]. Available:
  \url{https://dl.acm.org/doi/10.1145/3357223.3362719}
\BIBentrySTDinterwordspacing

\bibitem{misra2021rubberband}
U.~Misra, R.~Liaw, L.~Dunlap, R.~Bhardwaj, K.~Kandasamy, J.~E. Gonzalez,
  I.~Stoica, and A.~Tumanov, ``Rubberband: cloud-based hyperparameter tuning,''
  in \emph{Proceedings of the Sixteenth European Conference on Computer
  Systems}, 2021, pp. 327--342.

\bibitem{li2022easyscale}
M.~Li, W.~Xiao, B.~Sun, H.~Zhao, H.~Yang, S.~Ren, Z.~Luan, X.~Jia, Y.~Liu,
  Y.~Li \emph{et~al.}, ``Easyscale: Accuracy-consistent elastic training for
  deep learning,'' \emph{arXiv preprint arXiv:2208.14228}, 2022.

\bibitem{qiao_litz_atc2018}
A.~Qiao, A.~Aghayev, W.~Yu, H.~Chen, Q.~Ho, G.~A. Gibson, and E.~P. Xing,
  ``Litz: Elastic framework for $\{$High-Performance$\}$ distributed machine
  learning,'' pp. 631--644, 2018.

\bibitem{narayanamurthy2013towards}
S.~Narayanamurthy, M.~Weimer, D.~Mahajan, T.~Condie, S.~Sellamanickam, and
  S.~S. Keerthi, ``Towards resource-elastic machine learning,'' in \emph{NIPS
  2013 BigLearn Workshop}, vol.~1, no. 2.1, 2013, pp. 2--3.

\bibitem{li_ease.ml:_2017}
T.~Li, J.~Zhong, J.~Liu, W.~Wu, and C.~Zhang,
  ``\BIBforeignlanguage{en}{Ease.ml: {Towards} {Multi}-tenant {Resource}
  {Sharing} for {Machine} {Learning} {Workloads}},''
  \emph{\BIBforeignlanguage{en}{arXiv:1708.07308}}.

\bibitem{gu2019tiresias}
J.~Gu, M.~Chowdhury, K.~G. Shin, Y.~Zhu, M.~Jeon, J.~Qian, H.~Liu, and C.~Guo,
  ``Tiresias: A $\{$GPU$\}$ cluster manager for distributed deep learning,'' in
  \emph{16th USENIX Symposium on Networked Systems Design and Implementation
  (NSDI 19)}, 2019, pp. 485--500.

\bibitem{le2020allox}
T.~N. Le, X.~Sun, M.~Chowdhury, and Z.~Liu, ``Allox: compute allocation in
  hybrid clusters,'' in \emph{Proceedings of the Fifteenth European Conference
  on Computer Systems}, 2020, pp. 1--16.

\bibitem{mohan2022looking}
J.~Mohan, A.~Phanishayee, J.~Kulkarni, and V.~Chidambaram, ``Looking beyond
  $\{$GPUs$\}$ for $\{$DNN$\}$ scheduling on $\{$Multi-Tenant$\}$ clusters,''
  in \emph{16th USENIX Symposium on Operating Systems Design and Implementation
  (OSDI 22)}, 2022, pp. 579--596.

\bibitem{yu_computation_2019}
H.~Yu and R.~Jin, ``\BIBforeignlanguage{en}{On the {Computation} and
  {Communication} {Complexity} of {Parallel} {SGD} with {Dynamic} {Batch}
  {Sizes} for {Stochastic} {Non}-{Convex} {Optimization}},''
  \emph{\BIBforeignlanguage{en}{arXiv:1905.04346}}, 2019.

\bibitem{zheng_cynthia:_2019}
H.~Zheng, F.~Xu, L.~Chen, Z.~Zhou, and F.~Liu,
  ``\BIBforeignlanguage{en}{Cynthia: {Cost}-{Efficient} {Cloud} {Resource}
  {Provisioning} for {Predictable} {Distributed} {Deep} {Neural} {Network}
  {Training}},'' in \emph{\BIBforeignlanguage{en}{48th {ICPP} 2019}}, pp.
  1--11.

\bibitem{yao_large_2020}
\BIBentryALTinterwordspacing
Z.~Yao, A.~Gholami, D.~Arfeen, R.~Liaw, J.~Gonzalez, K.~Keutzer, and
  M.~Mahoney, ``\BIBforeignlanguage{en}{Large batch size training of neural
  networks with adversarial training and second-order information},''
  \emph{\BIBforeignlanguage{en}{arXiv:1810.01021 [cs, math, stat]}}, Jan. 2020,
  arXiv: 1810.01021. [Online]. Available: \url{http://arxiv.org/abs/1810.01021}
\BIBentrySTDinterwordspacing

\end{thebibliography}

\end{document}